\documentclass[structabstract]{aa}
\usepackage{graphicx,amsmath,natbib}
\usepackage[dvips]{changebar}
\usepackage{txfonts}
\newcommand\msol{\ensuremath{\,\mbox{\it M}_{\odot}}}

\newcommand\lta{\mathrel{\hbox{\raise 0.6 ex \hbox{$<$}\kern
                   -1.8 ex\lower .5 ex\hbox{$\sim$}}}}
\newcommand\gta{\mathrel{\hbox{\raise 0.6 ex \hbox{$>$}\kern
                   -1.7 ex\lower .5 ex\hbox{$\sim$}}}}
 
\newcommand{\scrbox}[1]{\ensuremath{{\mbox{\scriptsize #1}}}}

\newcommand{\teff}{{\ensuremath{T_{\scrbox{eff}}}}}
\newcommand{\Msol}{\ensuremath{\,\mbox{\it M}_{\odot}}}
\newcommand{\Mloss}{\ensuremath{\,\mbox{\it M}_{\odot}{\rm yr}^{-1}}}
\newcommand{\Mstar}{\ensuremath{\it \,M_{*}}}

\newcommand{\MS}{main--sequence}
\newcommand{\gr}{\ensuremath{g_{\scrbox{rad}}}}

\newcommand{\DM}{\ensuremath{ \log \Delta M/M_{*}}}
\newcommand{\Dm}{\ensuremath{ \Delta M/M_{*}}}

\newcommand{\He}{\mbox{He}}
\newcommand{\Be}{\mbox{Be}}

\newcommand{\Fe}{\mbox{Fe}}

\newcommand{\Ca}{\mbox{Ca}}
\newcommand{\Ni}{\mbox{Ni}}

\newcommand{\Ti}{\mbox{Ti}}

\newcommand{\Li}{\mbox{Li}}

\begin{document}

\title{AmFm and lithium gap stars}
\subtitle{Stellar evolution models with mass loss}
\author{M. Vick\inst{1,}\inst{3}, G. Michaud\inst{2,}\inst{3}, J. Richer\inst{3}, 
 \and O. Richard\inst{1}}


\institute{GRAAL UMR5024, Universit\'e Montpellier II,
                 CC072, Place E. Bataillon,
                 34095\,Montpellier Cedex 05,
                 France		
\and
            LUTH, Observatoire de Paris,
	    CNRS, Universit\'e Paris Diderot,
	    5 Place Jules Janssen, 92190\,Meudon, France           
\and
	   D\'epartement de physique, Universit\'e de Montr\'eal, 
           Montr\'eal, Qu\'ebec, H3C 3J7, Canada\\
           \email{mathieu.vick@umontreal.ca, michaudg@astro.umontreal.ca,\\
                  jacques.richer@umontreal.ca, olivier.richard@graal.univ-montp2.fr}
}

\date{Received February 24, 2010; accepted June 13, 2010}

\abstract{}
{A thorough study of the effects of mass loss on internal 
and surface abundances of A and F stars
is carried out in order to constrain mass loss rates for these stars, as well 
as further elucidate some of 
the processes which compete with atomic diffusion.}
{Self-consistent stellar evolution models 
of 1.3 to 2.5\,\Msol{} stars including atomic diffusion 
and radiative accelerations 
for all species within the OPAL opacity database were computed with 
mass loss and compared to observations 
as well as previous
calculations with turbulent mixing.}
{Models with unseparated mass loss rates between $5\times10^{-14}$ and 
$10^{-13}\Mloss{}$ reproduce observations 
for many cluster AmFm stars as well as Sirius\,A and $o$\,Leonis. 
These models also explain cool Fm stars, but not  
the Hyades lithium gap. Like
turbulent mixing, these mass loss rates reduce surface abundance 
anomalies; however, their 
effects 
are very different with respect to internal abundances. For 
most of the \MS{} lifetime
of an A or F star, surface abundances 
in the presence of such mass loss depend on separation which 
takes place between $\DM=-6$ and $-5$.}
{The current observational constraints do not allow us to conclude that mass loss 
is to be preferred over turbulent mixing (induced by rotation or otherwise) in 
order to explain 
the AmFm phenomenon. Internal concentration variations which could be detectable
through asteroseismic tests should provide further information. 
If atomic diffusion coupled with mass loss are
to explain the Hyades Li gap, the wind would need to be separated.}


\maketitle

\section{Astrophysical context}
\label{sec:intro}
As instruments become more sophisticated and precise observations 
are readily made available for a growing 
number of chemical elements, 
additional constraints are steering the 
evolution of stellar models. The inclusion of atomic diffusion and 
radiative accelerations into the standard 
stellar evolution model 
resulted in some early success in describing abundance patterns 
for AmFm stars \citep{turcotte98fm}. 
In these models, particle transport within the radiative zone 
was calculated from all physics known
through first principles. However, the calculated surface 
anomalies were greater than the 
observed anomalies, and 
it was 
determined that additional 
transport processes were flattening the surface abundance patterns. 
With the addition of turbulent 
mixing, the Montreal group's stellar evolution 
code \citep{richer00,richard01,michaud04,michaud05,talon06} was able to 
explain and reproduce the
particular abundances of many cluster and field stars with a single tunable 
parameter: the mixed mass. Nonetheless, 
other studies have suggested that mass loss could also reduce 
the predicted anomalies to observed levels \citep{michaud83,michaud86,alecian96}. 
Since these studies considered 
static stellar models incorporating a 
limited number of species, a rigorous investigation of the effects of 
mass loss on stellar chemical evolution is warranted. Previously, 
\citet{vauclair95} have introduced mass
loss to reduce the effect of Li settling in Pop\,II evolutionary models.

AmFm stars (7000\,$\leq$\,\teff\,$\leq$\,10\,000\,K) are slowly rotating 
($v_{{\rm rot}}<100$\,km\,s$^{-1}$, \citealt{abt00}), 
non-magnetic stars of the 
\MS{} (MS). They
are interesting candidates for testing evolutionary models because 
they lie within the temperature range for which the depth of 
the surface convection zone varies 
quite rapidly. Chemical separation within the stable radiative 
zones of these stars generates
surface abundance 
anomalies within timescales that depend strongly on the depth of the surface 
convection zone, although the exact depth at
which the separation occurs is 
still debated. The original explanation for these chemically peculiar stars 
stipulated 
that the separation
occurred immediately below the surface H convection 
zone \citep{watson71,alecian86,alecian96}; 
however, with the 
inclusion of turbulent mixing, more recent models
\citep{richer00} have suggested that separation 
occurs much deeper in the star ($T$\,$>$\,200\,000\,K). Though the second scenario
has had success not only with AmFm stars, but also with Pop\,II \citep{richard05,korn06} 
and Horizontal Branch (HB) stars
\citep{michaud07,michaud08}, it is 
still premature to accept turbulence as the sought-after macrospcopic 
process since all models with turbulence necessarily involve at least one
adjustable parameter. Turbulence models often implicitely assume a 
link between turbulence and
rotation; however, even the most slowly rotating AmFm stars have anomalies
which are significantly smaller than those obtained with atomic diffusion only models. 
This 
suggests the presence of a competing process even in non rotating stars.    

Typical anomalies on the surface of AmFm stars include overabundances 
of iron peak elements, as well as underabundances 
of Ca and/or Sc (see \citealt{cayrel91} for a more complete description). Recent 
studies have obtained
abundance determinations for numerous A and F stars of open clusters 
for a number of  
chemical species 
\citep{burkhart00,hui-bon-hoa00,monier05,fossati07,gebran08coma,gebran08pleiades} 
in an effort to confront relatively well constrained stars to current evolutionary models. 
The advantage of observing 
cluster stars is that they generally have the same age and initial metallicity, 
which greatly facilitates a comparison with models. 

At the cool end of the Fm star domain, 
the well documented Li dip, first observed in the 
Hyades open cluster \citep{boesgaard86}, 
has challenged theoretical astronomers for decades \citep{michaud86li,vauclair88,michaud91,
talon05}. 
The lithium abundance has also been observed in 
many other open clusters (e.g. \citealt{balachandran95,burkhart00,twarog09}).
The Be abundances 
in these stars \citep{boesgaard02,boesgaard04,randich07}  
provide additional constraints on particle transport. Recently, \citet{talon03,talon05} 
have modeled shear turbulence induced by differential 
rotation and mixing induced by internal gravity waves in order
to describe both the hot and cold side of the dip. Other models have also
explored the effects of horizontal $\mu$-gradients on rotationally 
induced mixing in these stars
(\citealt{palacios03}).
However, 
the potential effects of atomic diffusion, more specifically 
of radiative accelerations, in competition with mass loss, were not fully investigated. 

At some level, mass 
loss is present in all stars, and it is important to quantify its 
effects on observed abundances.
Unfortunately, for A and F stars in particular, the mass loss rates are not  
known. In O and B stars, the radiatively driven winds produce 
mass loss rates as important 
as $10^{-4}$\Mloss (\citealt{lamers99}). 
In colder stars of types G and K, 
winds are driven by active coronas. The best known example is 
the solar wind which has a   
mass loss rate of 2\,$\times$\,10$^{-14}$\Mloss\,(\citealt{feldman77}). 
For intermediate stars, 
our understanding is at best nebulous. \citet{abbott82} suggests that 
radiative accelerations are too small in stars with
\teff\,$\leq 10\,000\,$K for radiatively driven mass loss to be significant. 
On the other hand, the thinning of the surface convection zone
in F and particularly in A stars might be too important for solar type winds 
to exist (\citealt{parker60}). Are winds of A and F stars driven by radiation, 
coronal heating, both or neither? What are the expected mass loss rates? Both of these 
questions remain unanswered, and answering one could shed light on the other. 

Observational constraints on A and F star mass loss rates are limited. 
\citet{lanz92} as well as \citet{brown90} gave an upper limit of $10^{-10}$\Mloss 
for main sequence A stars. Asymmetries in  Mg\,II spectral lines of Sirius\,A led 
\citet{bertin95} to conclude that mass loss is present with a rate between  
$10^{-13}$\Mloss{} and $10^{-12}$\Mloss{}. On the theoretical side, 
the radiatively driven wind model of \citet{babel95} suggests a mass loss rate 
of $10^{-16}$\Mloss{} for A stars. However, according to his results, only 
heavier elements are evacuated by the radiative field. Similarly, 
\citet[see also \citealt{michaud86}]{michaud83} suggested that mass loss rates 
between $10^{-14}$\Mloss{} and $10^{-15}$\Mloss{} could 
satisfy observational constraints from CP star surface anomalies. Given the 
large disparity in 
values, the mass loss rates used in this study will be constrained 
stricly by surface abundance variations (see also Sect.\,4.2). 

In this paper, we consider mass loss in \emph{non rotating stars}. In these stars,
mass loss is arguably the only
macroscopic process competing with atomic diffusion within the radiative zones. 
We will start with a brief description of our stellar evolution 
code in Sect.\,\ref{sec:calcul} 
after which we will discuss the method for calculating radiative accelerations 
for lithium, beryllium 
and boron (Sect.\,\ref{sec:gradli}). In Sect.\,\ref{sec:mloss} we will describe the 
treatment of mass loss. 
In Sect.\,5 we will discuss its effect on internal structure and surface 
abundances as the models move 
along the \MS{} and the subgiant branch. 
In Sects.\,6 and 7 we will compare our models to
turbulence models and observations respectively. In Sect.\,\ref{sec:conclusions}, 
a brief overview of the main results will be followed by a discussion on how 
asteroseismology could help decipher the effects of advective 
processes including meridional circulation
from those 
incurred through turbulent processes caused by differential rotation.

\section{Calculations}
\label{sec:calcul}

This paper is part of the Montreal stellar model development project 
(\citealt{richard01}, \citealt{turcotte98soleil} 
and references therein). The models were evolved from
the initially chemically homogeneous pre--\MS{}
\begin{figure}[!t]
\begin{center}
\includegraphics[scale=.505]{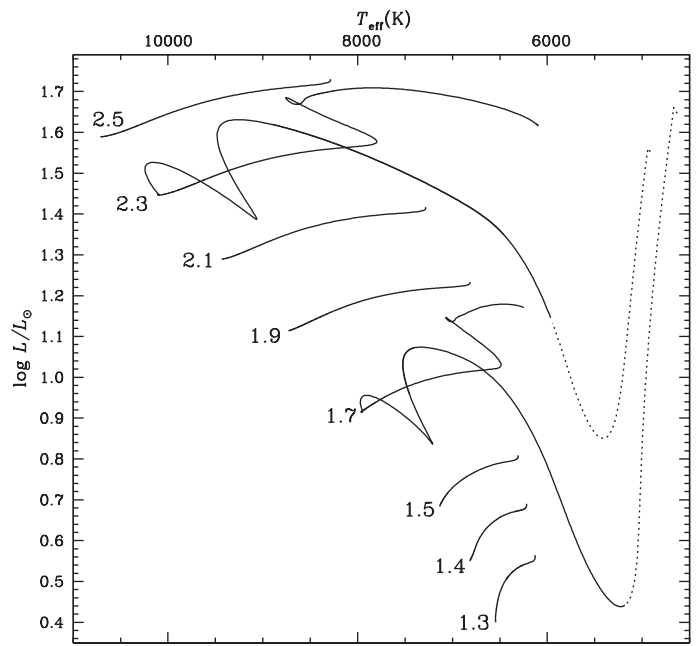}
\caption{H-R diagram for all the models shown in Fig.\,\ref{fig:hist}. 
Though all models were calculated from the PMS to the bottom of the subgiant branch, 
the complete tracks are only shown for the 1.7 and 2.3\,\Msol{} models. 
All other models are shown only on the \MS, which, for the purpose of this plot, is considered 
to span from the time at which 1\% of core H 
is burned to the time at which 95\% is burned. The dotted line represents evolution on the PMS
until diffusion starts (i.e. until a radiative zone appears). 
}\label{fig:HR}
\end{center}
\end{figure}
(see Fig.\,\ref{fig:HR}) with the solar abundance mix listed in Table 1 of
\citet{turcotte98soleil}. The transport of chemical elements in 1-D (one dimension) 
is solved within the basic 
framework established by \citet{burgers69}. The chemical transport 
equation considers 
all effects of radiative accelerations and atomic diffusion, and it 
is solved for 28 chemical 
elements and isotopes at every mesh point (the number of points varies from 800 to 
about 2800)
for each time step. Atomic diffusion is allowed to operate as soon as a 
radiative zone appears. 
The radiative acelerations (\gr) and the Rosseland mean opacity are
continuously updated; 
the treatment of chemical transport is thereby fully self-consistent. 
The atomic diffusion coefficients were taken
from \citet[see also \citealt{michaud93}]{paquette86}. The 
Krishna Swamy $T$--$\tau$ relation \citep{krishna66} was imposed as the 
boundary condition
in the atmosphere (this choice was motivated by results presented in
\citealt{vandenberg08}). Semiconvection was included as
described in \citet{richard01}, following \citet{kato66},
\citet{langer85}, and \citet{maeder97}.  

The chosen values of the mixing length parameter and initial He fraction are 
respectively $\alpha=2.096$ and  $Y_0=0.27769$ (see Model 
H of \citealt{turcotte98soleil}), which are 
calibrated by fitting the current solar radius and luminosity. 
We chose $Z_0=0.01999$ as the initial mass fraction of 
metals\footnote{ \citet{asplund05,asplund09} 
have proposed a downward
revision of the solar abundances of some metals; however we have 
chosen to keep the previous 
abundances until their 
determinations are reconciled  with helioseismology (\citealt{delahaye06,basu07} and 
references therein). The abundance of O, the third most abundant 
element, is particularly uncertain \citep{caffau08,delahaye10}.} .
Some models were also calculated for $Z_0=0.01$ and $Z_0=0.03$. 

These are the first fully self-consistent stellar models which include mass loss. 
Models were calculated from 1.30\,\Msol{} 
to 2.50\,\Msol. The mass loss rates considered vary 
from 1\,$\times$\,10$^{-16}$\Mloss{} to 1\,$\times$\,10$^{-12}$\Mloss. 
Our treatment of mass loss will be further discussed in Sect.\,\ref{sec:massloss}.

\section{Radiative accelerations}
\label{sec:gradli}

Rosseland opacities as well as radiative accelerations are 
continuously computed for all 28 species as the relative concentration of 
each species varies with time. 
For all elements included in the OPAL database \citep{iglesias96}, the 
radiative acceleration calculations are carried out 
using direct summations over the actual spectrum 
(i.e. $opacity\,\,sampling$, see \citealt{richer98}). 
At large optical depths (where the diffusion approximation is valid), 
the radiative acceleration, \gr($A$), of an element $A$ at 
a radius $r$ in a star may be approximated by:

\begin{equation}   
\gr(A)=\frac{1}{4\pi r^2}\frac{L^{\rm rad}_r}{c}\frac{\kappa_R}{X_A}\int_{0}^{\infty}\frac{\kappa_{u}(A)}{\kappa_u(total)}\mathcal{P}(u)du,
\label{eqn:grad}
\end{equation} 
where $\mathcal{P}(u)$, the normalized black body flux distribution, is given by:

\begin{equation}
\mathcal{P}(u)\equiv \frac{15}{4\pi^4}\frac{u^4e^u}{(e^u-1)^2}.
\end{equation} 
The radiative luminosity at a radius $r$ is $L^{\rm rad}_r$, $X_A$ is the 
mass fraction of $A$, $\kappa_R$ is the Rosseland 
opacity, $\kappa_u$(total) and $\kappa_u$($A$) are respectively the total 
opacity and the contribution of $A$ to
the total opacity at the frequency $u$ defined by:
\begin{equation}
u \equiv h\nu/kT. 
\end{equation}
Since the competition for photons between element $A$ and all other 
elements present in the plasma
determines the value of \gr($A$), atomic data is required for all 
species in order to compute \gr\, for any given 
element.     

The corrections for redistribution of momentum are 
from \citet{gonzalez95} and \citet{leblanc00}.

\subsection{Radiative accelerations for Li, Be and B}
\subsubsection{Methods}
\label{sec:methods}
Li, Be, and B are not included in OPAL since they are not significant in the 
calculation of Rosseland mean opacity\footnote{In \citet{turcotte98soleil} 
and \citet{turcotte98fm},
     \gr(\Li) and \gr(\Be) were approximated using formulae
     derived previously by \citet{richer93}.  These are
     however less accurate than \gr\, 
     calculated using direct summations over the spectrum
     throughout stellar evolution (using Eq.\,[\ref{eqn:grad}]),
    which are described here and were used in \citet{richer00} and \citet{richard01}.}.
We have nonetheless computed the radiative accelerations in a manner which is consistent with OPAL spectra.  
The various corrections determined by
\cite{richer97} are also included.
The atomic data required for LiBeB are taken from \citet{wiese66}. Since
these elements are not sufficiently abundant for pressure broadening to be important 
(the lines are never saturated), only oscillator strengths are required to compute opacities.  

The calculation of \gr\ for these elements is delicate because of two factors 
which can lead to large fluctuations: (1) non optimal frequency sampling and 
(2) random background changes.
Both problems arise from the fact that LiBeB not only have very few lines 
contributing to their 
\gr, but these lines are also very narrow. 

\begin{figure*}[!t]
\begin{center}
\includegraphics[scale=.68]{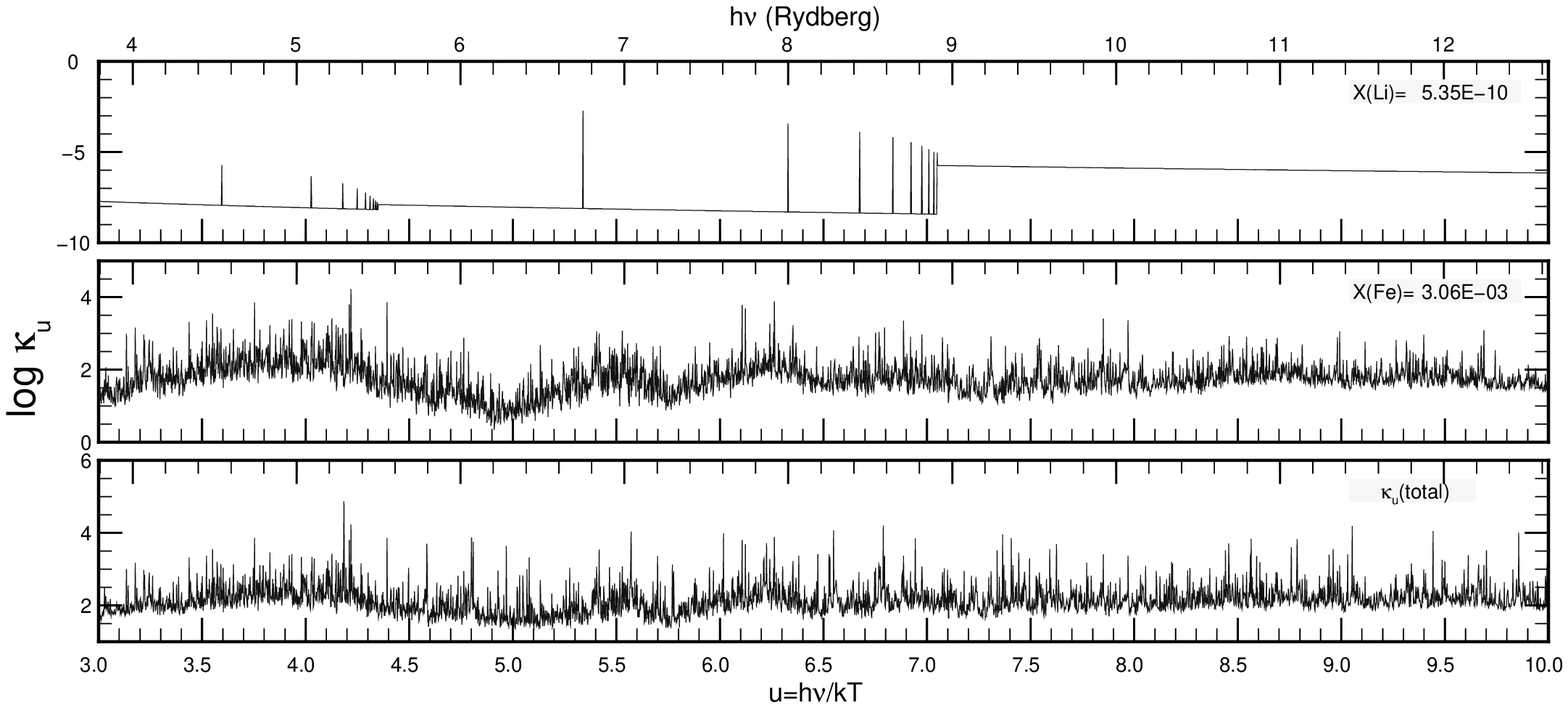}
\includegraphics[scale=.68]{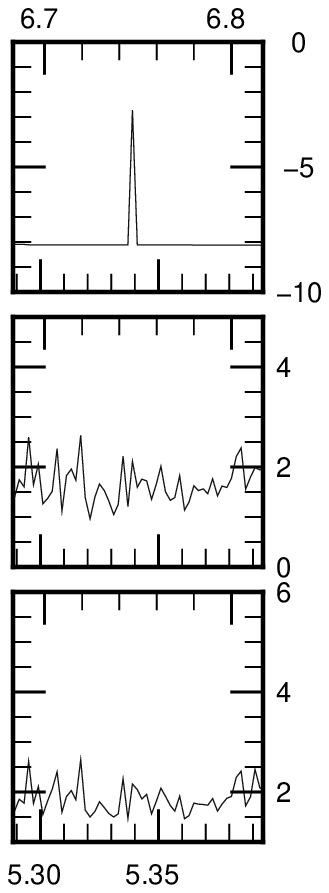}
\caption{Opacity spectra for Li, Fe as well as the total opacity 
in cm$^2$g$^{-1}$ at a depth where $\log T=5.3$ 
and $\log \rho=-4.6$. The $u$ interval covers the range through which most of 
the flux passes. The spectrum is dominated by Fe lines. The inset on the right shows a
zoom of the area which hosts the Li line which contributes the most to its \gr.
}\label{fig:spectre}
\end{center}
\end{figure*}

(1) The fluctuations due to sampling not being sufficiently refined
has an effect on Li as well as on the background (see Fig.\,\ref{fig:spectre}). The Doppler width 
of a $^7\Li$ line is given by:     
\begin{equation}
	\frac{\Delta u}{u} = \left[\frac{2\,k T}{Mc^2}\right]^{0.5} = 5.1 \times 10^{-5} T_5^{0.5}
	\label{eq:Doppler}
\end{equation}
          where $u$ is the adimensional energy difference between the upper and 
	  lower levels of the transition, $M$ is the mass of the element, $T$ is the local
	  temperature
          and $T_5=T/10^{5}$\,K.
          In this case, one 
	  typically has $u \simeq 5$  around the most important temperature range, 
	  $\log T \simeq 5.3$, so that          
\begin{equation}
	\Delta u_{\mathrm{Doppler}} = \Delta u_D \simeq 7 \times 10^{-5}
	\label{eq:deltadoppler}
\end{equation}
which is to be compared to the frequency grid interval, 
$\Delta u_G = 20/10^4 = 2 \times 10^{-3}$, of the OPAL spectra.     
Using opacity sampling for \Li{} would imply that the line center would 
sometimes occur \emph{within} $\Delta u_D$ of a grid
point and sometimes up to $15 \Delta u_D$ away. If the sampling point 
occurs in the far
tail of an important Li line, 
the \gr(Li) value would be much smaller 
than if it occurs in its center. 

Likewise, the sampling is not refined enough to reproduce all features of 
the background spectra (see Fig.\,3 of \citealt{richer98}). Variations can therefore
result from overlooking an important background line which occurs near a lithium line.

(2) The frequency sampling grid is a function of $u$ and not of $\nu$. 
Therefore, when one considers an adjacent grid point of different $T$, 
the background, as a function of $u$, changes, and 
the consequent random variation in background affects 
the flux available for a specific lithium line. This is 
largely due to the narrowness of most lines. For example, 
in the inset of Fig.\,\ref{fig:spectre}, the Li line does not encounter any important Fe line
for that specific ($\rho$,$T$) table; however, in the inset of 
Fig.\,1 of \citet{richer05}, which
shows the same $u$ interval for a different ($\rho$,$T$) table, 
the Li line is overlapped by a strong Fe line, and the available flux is consequently reduced. 

One can alleviate these problems by combining a modified version of the opacity 
sampling method with the known position of all 
the line centers for LiBeB. In order to reduce the
fluctuations, while preparing the spectra of, say, \Li{}, for the
calculations of \gr(Li), it was assumed that each \Li{} line was
spread uniformly over the $\Delta{}u_G$ interval in which the \Li{} line
center occurs (i.e. the Li lines become square functions of width $\Delta{}u_G$ and so
it becomes impossible for the sampling to overlook them).  This will leave fluctuations caused by
random variations of the opacity background when the line
center moves from one $\Delta{}u_G$ interval to another (2), but the variations due to
non optimal sampling (1) will be reduced significantly. 

Note that if the frequency sampling is refined (e.g. 10$^5$ as
in OP data) in order to better represent the background spectra, 
errors due to the inexact 
position (in frequency) of each background feature remain.
The line 
center positions for Fe, the main contributor 
to the total opacity in this region, are only known to about 1\%. 

The resulting error bars on \gr{} are discussed in Sect.\ 3 of \citet{richer05}. 
As an example, for population I stars, there is a factor of 2 uncertainty 
for \gr(Li). In Pop\,II stars, the uncertainty is much less important 
(e.g. only 3\% for a star with $Z=0.0001$).  

The same errors and limitations should be expected for scandium 
around the minima of its \gr{} since the radiative
accelerations in these regions are computed with only a few narrow lines. Unfortunately, Sc 
is not included in OPAL data, and its \gr{} must be calculated through 
alternate methods (\citealt{leblanc08}).

\begin{figure}[!t]
\begin{center}
\includegraphics[scale=.48]{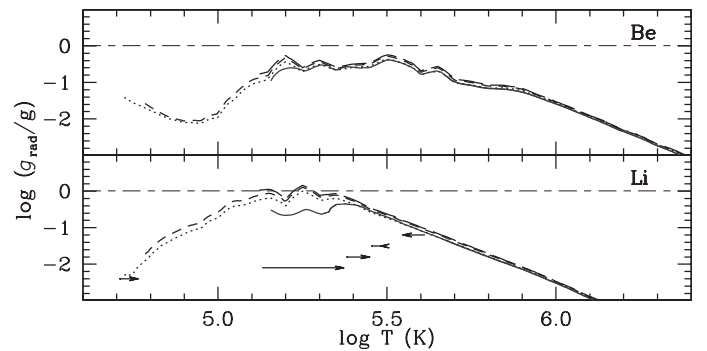}           
\caption{Variation of radiative accelerations with temperature for Li 
and Be at 100\,Myr 
(dotted line), 
700\,Myr (dashed line) and 1.3\,Gyr (long dashed line) in a 
1.55\,\Msol{} model with a mass loss rate of
$1\times10^{-13}\Mloss$ as well as at 100\,Myr (solid line) for a 1.47\,\Msol\ 
model with no mass loss. 
The curves extend to the bottom of the surface convection zone, on the left.
Acceleration is normalized to local gravity. 
The horizontal arrows indicate the total movement of the bottom of the
surface convection zone (BSCZ) for models of 1.40, 1.43, 1.45, 1.47 and 1.55\,\Msol{} 
from top to bottom (see text below). 
}\label{fig:gradLiBe}
\end{center}
\end{figure}

\subsubsection{Results}
\label{sec:resultsLiBeB}
The radiative accelerations for Li and 
Be are shown in Fig.\,\ref{fig:gradLiBe} for 
a star of 1.55\,\Msol{} with 
a mass loss rate of $1\times10^{-13}\Mloss$ as well as for 
a star of 1.47\,\Msol{} with no mass loss. 
For both models, the radiative accelerations for Be are below
gravity throughout the radiative zone, and throughout evolution. 
However, \gr(Li) has a peak that approximately reaches 
gravity below the surface 
convection zone for the 1.55\,\Msol{} model. 
Up to $\log T\simeq 5.4$, \gr{} is seen to 
be almost solely temperature dependant as the curves for both models overlap. 

In the 1.47\,\Msol{} model, \gr(Li) has about the same value as in the 1.55\,\Msol{} 
model for $\log T > 5.4$, but, for $\log T < 5.4$ it is reduced
by the competition of Fe, which absorbs most of the 
flux where the Li lines are important. This occurs when
mass loss is small enough for Fe to accumulate 
in and above this region, as is the case for the
1.47\,\Msol{} model (or in the 1.50\,\Msol{} model with a mass loss rate of
$10^{-14} \Mloss$ shown in Fig.\,\ref{fig:abint1.5}).  
On the other hand, \gr(Be) does not vary with Fe abundance likely because
most Be
lines, particularly the ones which contribute the most to its \gr, 
lie outside of the
frequency interval which is dominated by Fe lines. 

The horizontal arrows in Fig.\,\ref{fig:gradLiBe} show 
the total movement of the bottom of
the surface convection zone (BSCZ) for 4 models without mass loss 
(1.40, 1.43, 1.45 and 1.47\,\Msol), as well as 
for a model of 1.55\,\Msol{} with a mass loss rate of 
$10^{-13}\Mloss$. The interval for which the BSCZ is plotted
spans from 50\,Myr to 625\,Myr for all models except for the 1.47\,\Msol, model
which goes from 50\,Myr to 425\,Myr (the simulation's last converged model). 
For most models, the depth of the 
BSCZ does not vary much between 50 and 625\,Myr (most of the \MS{} lifetime before
the age of the Hyades cluster); however, for the 1.47\,\Msol{} model, the 
BSCZ moves significantly and extends over an interval in which \gr(Li) also 
varies significantly. 


If we compare the curves for the 1.55\,\Msol\ model with those shown in 
Fig.\,1 of \citet{richer93}, we first note the 
similarity in the temperature dependence
of the curves. For Li, we have verified that they are nearly equal for $\log T > 5.4$. 
The maxima occur at very nearly the same temperature as well. 
However, the maximum values of \gr(Li) and \gr(Be) are respectively 
about 2 and 6 times larger in \hbox{\citet{richer93}}. Furthermore, 
their results show very smooth curves
compared to the many variations seen in our calculations. 

Although there is a slight difference in stellar mass
(a 1.54\,\Msol{} model is shown in Fig.\,1 of \citealt{richer93}), 
this cannot account for the relatively large differences close to
the maximum. 
There are nonetheless a few other explanations. 
Our calculations 
were carried out with integrations over
complete OPAL spectra for all species at each time step and each mesh point. 
Since these spectra were not available in
1993, the calculations of 
\citet{richer93} were done using averaged spectra whose frequency
dependence (Eq. [18] of \citealt{borsenberger79}) did not 
include the frequency dependence of
Fe lines, particularly 
near $T \simeq 2 \times 10^5$\,K, where \gr(Li) is affected the most. 
From Fig.\,\ref{fig:spectre}, one can see that Fe lines dominate
the region where Li lines are strongest. As Fe lines which occur near Li 
lines 
were absent in the old opacities, the available 
flux for Li lines is very
different (see Sect.\,\ref{sec:methods}).    

Moreover, the evolution of individual metal abundances and its impact on 
local opacity are $not$ included in 
the calculations of \citet{richer93}. This is particularly 
important in AmFm stars since heavier metals such as iron and nickel 
tend to accumulate below
the surface convection zone when mass loss is not too strong.



\section{Mass loss}
\label{sec:mloss}
We first discuss how to include mass loss
in evolutionary models (Sect.\,\ref{sec:massloss}). In Sect.\,4.2,
we discuss a few theoretical analyses of coronal and radiatively driven winds followed 
by a brief look into separated winds\footnote{In this paper, mass loss is $separated$ 
when the abundances in the wind are not the same as in the photosphere. 
On the contratry, mass loss is $unseparated$
when the abundances in the wind are the same
as in the photosphere.} (Sect.\,4.3).
\subsection{Treatment of mass loss}
\label{sec:massloss}
Mass loss is assumed spherical and chemically unseparated.  
If we simply apply mass conservation arguments, the net result of 
mass loss is the appearance of 
an outward flowing interior velocity due to the wind:
\begin{equation}
v_{w}(r)= -\frac{\dot{M}}{4\pi r^{2}\rho}\frac{m_r}{M_*}
\label{eqn:vwind}
\end{equation}
where $\rho$ is the local density, $m_r$ is the mass interior to $r$ and $\dot{M}<0$. 
It should be pointed out that the 
$m_r/M_*$ factor in this equation has often been overlooked, but it will be shown
to be a consequence of mass conservation (see Sect.\,\ref{sec:demonstration}). 
Even for small mass loss rates, the $\rho^{-1}$ dependence leads to 
large velocities in the outer layers of the star. To avoid numerical 
instabilities due to the cancellation of two large quantities in the 
convective envelope (wind velocities and large turbulent velocities used 
to enforce convective mixing) and in order to have a Neumann surface 
boundary condition (no flux),  
mass loss was implemented as described in \citet{charbonneau93}. The main 
physical considerations are that the surface convection zone (SCZ) be 
fully mixed, and that the atmosphere be mixed with it by overshooting. 
Therefore, the ejected mass has the same composition as the SCZ. 
The mechanism by which mass is ejected from the star becomes irrelevant 
so long as the correct amount of mass is removed from the star. 
For each chemical species, the resulting transport equation is:
\begin{eqnarray}
\nonumber\rho\frac{\partial c}{\partial t}
&=&-\nabla\cdot\mbox{[}-\rho D{\bf\nabla}\ln c+\rho({\bf U}+{\bf U}_{w})c\mbox{]}\\
&&+\rho (S_{nuc}+S_{w})c,
\label{eqn:charb2}
\end{eqnarray}
with a Neumann condition imposed at the surface and with ${\bf U}_{w}$ and $S_{w}$ defined as:    
\begin{equation}
{\bf U}_{w}=\left\{\begin{array}{l}v_{w}{\bf\hat e_{r}}$ $ \mbox{under the SCZ,}\\ 0$ $ \mbox{in the SCZ;}\end{array}\right. 
\label{uw}
\end{equation}
\begin{equation}
S_{w}=\left\{\begin{array}{l}0$ $ \mbox{under the SCZ,}\\ \frac{\dot M}{M_{{\rm CZ}}}$ $ \mbox{in the SCZ.}\end{array}\right.
\label{eqn:sw}
\end{equation}
Here, $c$ is the time and depth dependent composition, 
$D$ the total diffusion coefficient, ${\bf U}$
the advective part of the atomic diffusion velocity, ${\bf U}_w$ 
the interior velocity due to the wind, $M_{{\rm CZ}}$ the mass of the SCZ, $\dot M$ the mass loss rate, 
$S_{nuc}$ a source/destruction term linked to nuclear reactions and 
$S_w$ is a sink term linked to mass loss. The mass is simply 
removed from the surface convection zone which, given the assumed mixing, is 
equivalent to losing it through the stellar surface. 
The mass of 
the star is also continuously updated so that all quantities 
that depend on stellar mass are correctly calculated.      
\subsubsection{Mass flux equation}
\label{sec:demonstration}
Mass loss has the effect of ejecting (or $peeling$) the outermost layers of a stellar model. 
The model must then be reconverged with a slightly reduced mass. This scenario can 
be described within the formalism of operator splitting. If a mass 
$\Delta M_{\rm{loss}}$ is removed in a given time step $\Delta t$, one needs to 
reconverge a model of mass
$M_1 = M_0 - \Delta M_{\rm{loss}}$ at time $t_1=t_0+\Delta t$ 
\textit{with the composition of the star kept unchanged as a 
function of 
$m_r$, the 
independent variable which is defined as the sphere of constant mass, $m_r$}. 
Mass loss does not modify the abundance profiles with respect to $m_r$, although 
there is a change with respect to the stellar surface. 
For example, an abundance peak at the point $M_0 -\Delta M_{\rm{loss}}$ 
(or $\Delta M_{\rm{loss}}$ below the surface before peeling), finds itself 
at $M_1$ (or at the surface) after the peeling but 
the mass interior to this point, $m_r$, has not changed.

To implement mass loss in a stellar evolution code, 
one can imagine the solution process to be broken down into 2 steps. 
In a first step, the model can be converged at a time $t_1$
as if there were no mass loss, with 
the composition changes due to nuclear reactions and diffusion processes. 
Then, the grid at $t_1$ is reinterpreted as
corresponding to a mass $M_1 = M_0 - \Delta M_{\rm{loss}}$
with all variables expressed as a function\footnote{Here $M_1 = M_*$ at $t_1$
and similarly $M_0 = M_*$ at $t_0$.} of $m_r/M_*$ except for the composition which 
is kept unchanged as a function of $m_r$. 
The star is then  reconverged 
a second time at $t_1$ with a mass $M_1$ and with the concentration
profiles which include the effect of mass loss. As more mass is lost, 
the model inches toward less massive structures, and so evolution 
progressively corresponds to that of a lower mass star. This is a 
correct description of mass loss which in practice 
is as accurate as the operator splitting
procedure is accurate.
 
However, going from the $m_r/M_0$ to the $m_r/M_*$ grid 
while keeping $c_i$ as a function of $m_r$ implies
interpolating. In practice, performing interpolations on $c_i$ 
during computations can introduce unwanted numerical diffusivity. 
It is relatively easy to show 
that these interpolations on $c_i$ resulting 
from mass loss can be avoided by introducing 
a local mass flux into the conservation equation, while keeping the grid constant (i.e. the grid points
have the same values of $m_r/M_*$ in spite of $M_*$ varying from $M_0$ to $M_1$).

Let us consider the concentration 
on a grid point ($m_r/M_*$) at $t_1$ as a function of the concentration 
on the same grid point
at $t_0$. The grid points shift on the $m_r$ axis.  Then 
\begin{equation}
c_i \left(m_{r1}\right) =  c_i \left({m_{r0}}\right) + \frac{\partial c_i}{\partial  {m_r}} {\Delta m_r}
        \label{eq:concen_shft}
\end{equation} 
which is a simple Taylor series development and where   
\begin{equation}
\Delta m_r = m_{r1} - m_{r0}.
        \label{eq:deltamr}
\end{equation} 
By definition, $m_{r1}$ is the position on the $m_r$ axis, of the 
grid point $\frac{m_{r0}}{M_0}$ after  a  $\Delta M_{\rm{loss}}$ mass loss
so that one may write:
\begin{equation}
\frac{m_{r1}}{M_1} = \frac{m_{r0} + \Delta m_r}{M_0 - \Delta M_{\rm{loss}}}= \frac{m_{r0}}{M_0}
        \label{eq:grid}
\end{equation}
which can trivially be solved to give:
\begin{equation}
 \Delta m_r= - \frac{m_{r0}}{M_0}({ \Delta M_{\rm{loss}}}).
        \label{eq:deltamr1}
\end{equation}
Now we substitute equation (\ref{eq:deltamr1}) in equation (\ref{eq:concen_shft}),  
replace $\Delta M_{\rm{loss}}$ by $-\dot{M}\times \Delta t$ and simplify:
\begin{equation}
c_i \left(m_{r1}\right) =  c_i \left(m_{r0}\right) + \frac{\partial c_i}
{\partial  {m_r}}\frac{{\rm d} \Mstar}{{\rm d} t } \frac{m_{r0}}{M_0}\Delta t.
        \label{eq:concen_shft1}
\end{equation} 
The second term on the right has the same effect on $c_i$ as the 
introduction of a flux term\footnote{Often abusively called a ``wind'' term in this context.  It seems 
preferable to restrict the use of ``wind'' to the region above the photosphere.} 
(caused by mass loss $\times \frac{m_r}{M_0}$) in a conservation equation. The flux $4 \pi \rho r^2 U_W$ given by equation 
\begin{equation}
 - 4 \pi \rho r^2 U_w = \frac{{\rm d} \Mstar}{{\rm d} t} \frac{m_r}{\Mstar}.
        \label{eq:masslossflux}
\end{equation}
must then be introduced in the conservation equation in order to take into 
account the effect of the $m_r$ shift caused by the peeling of surface layers 
while keeping the same grid as a function of $m_r/\Mstar$ when \Mstar{} changes 
due to mass loss. This simple argument justifies 
the introduction of a mass flux to model the effect of 
mass loss on element separation in stellar atmospheres 
or envelopes \citep{vauclair75,michaud83,alecian96}.

For smaller mass loss rates, structural effects of mass loss are often negligible 
to the extent that only the shift of $c_i$ remains 
and mass loss may be viewed as a mass flux going through a star of constant mass (i.e. it is not necessary to change the stellar mass during evolution calculations).  
Consider the case of the present Sun.  
Assuming that it is constant in time, its current mass loss rate of $2 \times 10^{-14}\Msol/$yr leads 
to a loss of $10^{-4}$\,\Msol{} up to the Sun's age which
for most purposes corresponds to negligible structural changes.
 This is only $\sim 1/300$ of the mass of the superficial convection zone.  

Since we are not certain about the nature of the winds at the surface of A and F stars, 
the present models do not take into account any energy dissipation which would be 
required to produce these winds. 



\subsection{Stellar winds of A and F stars}
This study will use observed surface abundances to constrain 
mass loss since the stellar winds associated with A and F stars are 
not well known. The winds could be radiative, coronal, a combination of both or even
completely negligeable. 
Comparisons with stellar wind models are difficult. Indeed, 
even the hottest A stars maintain a thin 
surface H convection zone, and stars as early as A7 (and possibly earlier, 
see \citealt{neff08,simon97}) can support 
active coronas and chromospheres which could harbor solar type winds. 
It is also plausible that both 
mechanisms act simultaneously. A few properties of coronal and 
radiatively driven winds for A and F stars will be
reviewed in the two following subsections, in so far as they relate to chemical separation. 

\subsubsection{Coronal winds}
\label{sec:coronal}
With simple physical considerations, it is possible to obtain an 
approximate value of the mass loss rate above which coronal winds 
are necessarily unseparated. For a spherically
symmetrical mass loss, the wind velocity (Eq. [\ref{eqn:vwind}] with $M_r=M_*$) may 
be compared 
to the maximal downward 
diffusion velocity (given by the gravitational settling
velocity without any contribution from radiative accelerations).
Equating the two gives an evaluation 
of the maximum mass loss that allows for separation to occur. 
One may then write: 
\begin{equation}
v_{w}= \frac{\frac{-{\rm d} M_*}{{\rm d}t}}{4\pi r^{2}\rho}=D_{ip}\frac{A_igm_p}{kT}.
\end{equation}
which can be rewritten as:
\begin{equation}
\frac{-1}{M_*}\frac{{\rm d} M_*}{{\rm d}t}=\frac{2.4 \times 10^{-15}A_iT_5^{1.5}}{Z_i^2}[{\rm yr}^{-1}]
\end{equation}
($T_5=T/10^5$\,K) where $A_i$ and $Z_i$ are the atomic mass and atomic number 
respectively of element $i$ and where the Coulomb term in the calculation of $D_{ip}$ is replaced by an approximate value. 
This applies both
in the atmosphere and in the outer parts of the wind solution. 
For a solar type wind, assuming an isothermal corona of $T=10^6\,$K, one obtains a
limiting mass loss rate $\simeq 10^{-13}$\Mloss (see Michaud et al. 1987). 
The dependence on $T$ is valid when the gas is
fully ionized. In the cooler atmospheres of early A stars (say at $10^4$\,K), 
He is mainly neutral, nevertheless the limiting mass
loss rate is approximately the same in the atmosphere as in the corona since the 
diffusion coefficients are larger by a factor of $200-300$ than those of ionized He (Michaud et al. 1978). 
The separation observed in
the solar wind ($\dot M \simeq 10^{-14}$\Mloss) agrees roughly with this analysis. 
As previously mentioned, observations show that elements 
which are neutral at $T=10^4\,$K  
are $\simeq 2-3$ times less abundant in the corona than in the 
photosphere \citep{meyer85,meyer96}.
Most agree 
that the separation occurs somewhere in the solar 
chromosphere immediately above the photosphere and could therefore
have an impact on observed photospheric abundances. For 
instance, \citet[in their model 4]{geiss86} 
find that matter which arrives at 1 AU has the same He to H ratio 
as at the wind's base.
Since there is no 
net separation within the solar wind, it is precarious to infer that 
all coronal type winds are chemically differentiated. Moreover, 
wind structures in A star 
coronas would probably be quite different 
since many metals that are neutral in G and F 
star photospheres are ionized in A star photospheres. 
It is thus very difficult to compare the solar wind 
to other coronal winds, particularly for hotter stars.     

\subsubsection{Radiatively driven winds}

\citet{babel95} found that for stars within $8000 \le \teff \le 14000\,$K, 
all radiatively driven winds must be fully separated. In these stars 
the Coulomb coupling is not sufficient to redistribute the momentum acquired 
by the heavier, radiatively accelerated elements onto the bound, more abundant 
H and He \citep{springmann92,owocki02,krticka03}. Since only metals are ejected from the star, 
the mass loss rates are much smaller: between $10^{-16}\Msol$ and 
$10^{-17}\Msol$. However, 
the multicomponent hydrodynamical model put forth in 
\citet{babel95} and \citet{babel96} only 
considers an average metal, rather than solving for each metal individually, 
therefore metal-specific mass loss rates are not known.
Furthermore, the interaction of radiatively driven winds 
with magnetic fields as well as convection is still poorly 
understood. This is particularly important for A stars, and the consequent 
uncertainties require us to be cautious before constraining our analysis with these results.  
For cooler stars, such as F stars, radiative accelerations are not believed to be able to 
generate significant mass loss \citep{abbott82}.

\citet{unglaub08} found 
that radiatively driven winds of sdB stars must be 
separated (accelerated metals cannot drag H and He) 
if the mass loss is smaller than $10^{-12}$\Mloss{} (10 times larger 
than the approximate value obtained in Sect.\,\ref{sec:coronal}). However, 
although 
the author's calculations are quite thorough, they are not complete 
in so far as some other poorly understood factors could play a significant 
role in determining wind properties. How does convection or magnetic fields, 
particularly flux tubes, affect the wind structure and velocity? Furthermore, 
the omission of line shadowing in the calculations could have a significant 
impact on the author's results\footnote{Line shadowing occurs when 
the wind velocity is not sufficient to Doppler shift the wind's line 
centers away from flux attenuated photospheric line centers, thus 
reducing radiative accelerations. This is often true until the 
wind reaches the sonic point.}. In fact, the author stipulates in Sect.\,6.2 of 
his article that including 
line shadowing could diminish \gr\, by a factor of 100 for the stronger 
photoshperic lines, thus leading to an overestimation of $\dot M$ 
by a factor of 10.

\subsection{Unseparated vs. separated mass loss}
\label{sec:regimes}
The object of this study is to 
constrain the effects of mass loss $solely$ via observed abundance anomalies. 
To do so, we use simple 
wind models in order to minimize the arbitrariness of the analysis. 
Accordingly, most calculations were done assuming simple unseparated winds, although 
a few calculations were also carried out assuming
separation in the wind in order to assess potential effects. 
Three cases of separated winds were considered: (1) 
only metals are ejected, (2) 
the separation mimics the solar wind with H treated as a high-FIP (First Ionization
Potential; \citealt{meyer85}) element and (3) 
the separation again mimics the solar wind but with H as a low-FIP element. 

In case 1, all metals are ejected with the same composition as the stellar surface, 
while H and He remain bound. For this scenario, mass loss rates were varied from 
$10^{-17}\,$to$\,5\times10^{-16}\Mloss$ in order to
account for the fact that only metals are leaving the star 
(around 2\% of the superficial mass fraction). 

The other scenarios (2 and 3) consider chemical 
separation in the solar wind as established by \citet{meyer85}, who found that elements with a 
FIP smaller than 9eV were 
approximately 4 times more abundant relative to hydrogen in the corona than in the 
photosphere, while higher-FIP elements, including hydrogen, kept their photospheric abundances. Although 
this scenario is generally favored, \citet{meyer96} questioned 
his own results a decade later by implying that instead of having overabundant low-FIP 
elements in the wind, higher-FIP
elements, including H, could be depleted 
in the corona\footnote{This question remains unanswered (see \citealt{feldman03} 
for a complete review).}. Both configurations are investigated: case 2 has H as a 
bound high-FIP element,
and case 3 has it as a low-FIP element.
Our approach was to divide all elements into two groups: low-FIP elements (below 11 eV) and high-FIP elements 
(He, C, N, O, Ne, Cl and Ar). All low-FIP elements were depleted 4 times 
faster than high-FIP elements.\footnote{Helium is assumed to have a 1:1 ratio with high-FIP elements, since the observations which
suggest that He would have 
a ratio of 1:4 with these elements are questioned (see \citealt{feldman03}).} 

Numerically, in cases (2) and (3), the destruction term in \hbox{Eq.\,[\ref{eqn:sw}]} was multiplied by a factor of 4 in the 
SCZ for all low-FIP
elements. 
A weight term, which was continuously updated as concentrations changed in the SCZ, was
added to the denominator for normally depleted high-FIP elements to account for the fact that
their destruction is not, in this case, proportional to the total mass loss rate multiplied by individual 
concentration (Eqs.\,[\ref{eqn:charb2}] and [\ref{eqn:sw}]) since their relative concentration
in the photosphere is not the same as in the wind. 
The interior wind velocities are not affected since the $v_{{\rm wind}}$ term does 
not depend on relative concentrations, but
simply on the mass loss rate.

\section{Evolutionary models}
\label{sec:EVOLUTION}
\begin{figure*}[!t]
\begin{center}
\includegraphics[scale=1.01]{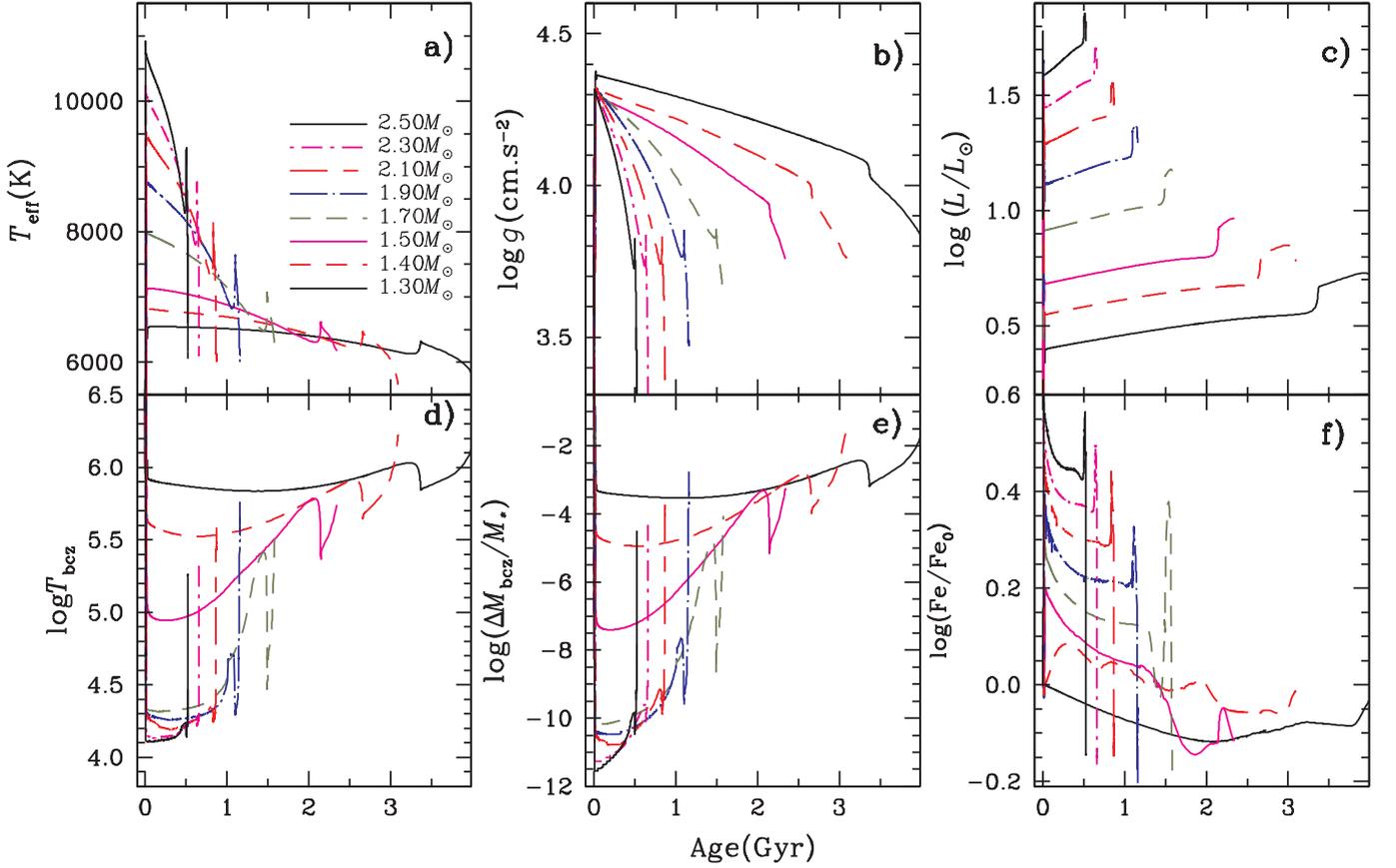}
\caption{Evolution of \teff{} (a), $\log g${} (b), luminosity (c), 
temperature  (d) and mass at the bottom of the surface convection 
zone (e), as well as the surface Fe abundance (f) for selected masses that span the range 
of AmFm stars. All models are shown for a mass loss rate of $10^{-13}$\Mloss. The curves are
identified in panel (a). Vertically, they have the same order in panels (a), (c) and (f), but 
are in reverse order in panels (b), (d) and (e).
}\label{fig:hist}
\end{center}
\end{figure*}

In Fig.\,\ref{fig:hist}, the evolution of \teff, $L$, $\log g$, the depth of the surface
convection zone as well as the surface Fe abundance are shown for a number of models. The chosen 
masses were selected to span the observed \teff\, range associated with 
AmFm stars and Li gap stars
(\citealt{preston74,boesgaard86}). The lower limit 
also corresponds to the least massive star on the main sequence for which 
we predict relatively large surface abundance anomalies (see Sect.\,\ref{sec:surfab}). 
All models are 
shown for the same mass loss rate ($10^{-13} \Mloss$). As will be seen in the 
next section, this 
mass loss rate has an effect, though very moderate, on the 
structure via abundance changes (the effect on He 
abundance for instance can increase the depth of the surface
convection zone). This mass loss rate also allows surface abundance anomalies
which are compatible with observed abundance anomalies 
for Am stars (see Sect.\,\ref{sec:surfab}). 

The \MS{} lifetime ranges from about 500\,Myr for the 2.5\,\Msol{}
model to more than 3\,Gyr for the 1.3\,\Msol{} model. 
The Fe surface abundance is intimately coupled with the 
movement of the surface convection zone; for the models 
with the thinnest SCZ, Fe is predicted to be overabundant 
over most of the MS lifetime (in the 2.5\,\Msol{} model 
by a factor of about 3). 

All models of at least 1.5\,\Msol{} are marked by a rapid Fe abundance 
peak which occurs at
the beginning of the main sequence. The rise of $X$(Fe), which is related to the 
depth change of the surface convection zone as the star stabilizes 
(see Fig.\,\ref{fig:hist}f), is so rapid that it is not 
resolved in Fig.\,\ref{fig:hist}f. Similarly, there is a sharp spike towards the end of the 
MS for most models which is once again
correlated with the sharp variation in SCZ depth 
(in this case the most important variation is for the
1.7\,\Msol{} model, which has a difference of a factor of 2.5 in Fe abundance 
within less than 100\,Myr). As we will see in
the following sections, these variations are larger when the mass loss rate is smaller. 
The HR diagram for these models is shown in Fig.\,\ref{fig:HR}. For clarity, 
only two evolutionary tracks are 
shown from the initially homogeneous 
pre-\MS{} up to the subgiant branch 
(the others are shown from the beginning of the \MS{}, although 
their evolution
is calculated from the pre-\MS{}). Diffusion and its effects 
on surface abundances appear well before the
arrival on the \MS{} (a thorough investigation of diffusion 
on the pre-\MS{} will be discussed in Vick et al., in preparation).      

\subsection{Radiative accelerations, internal abundance variations and structure}
\label{sec:gr_ab}
\begin{figure*}[!t]
\begin{center}
\includegraphics[scale=1.01]{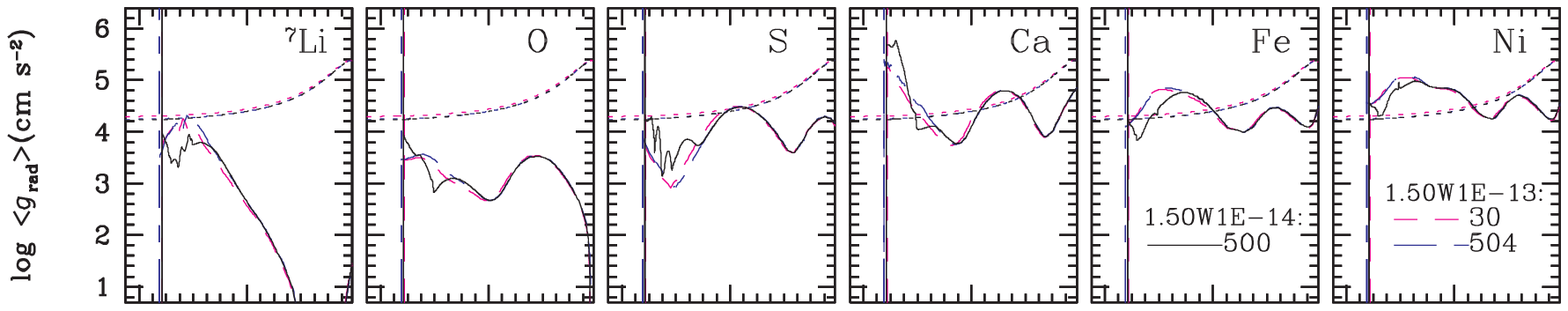}
\includegraphics[scale=1.01]{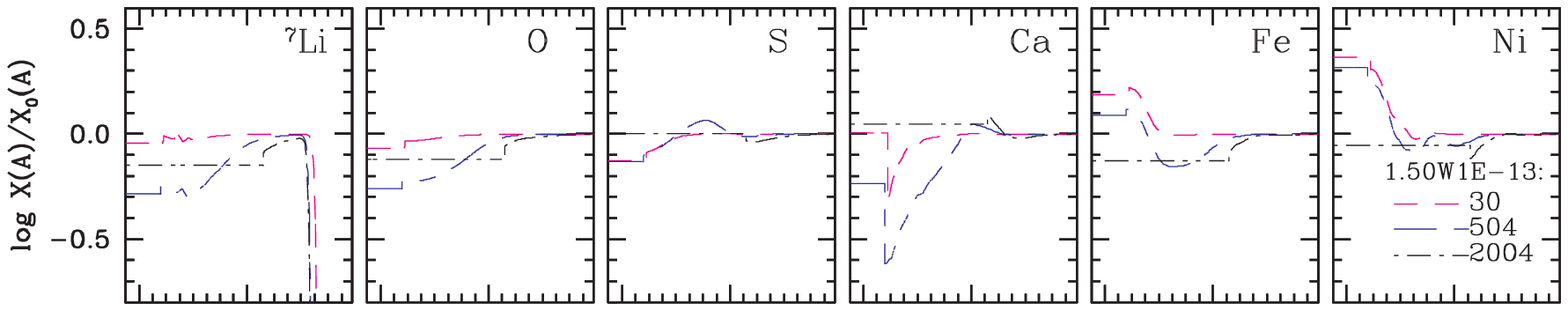}
\includegraphics[scale=1.01]{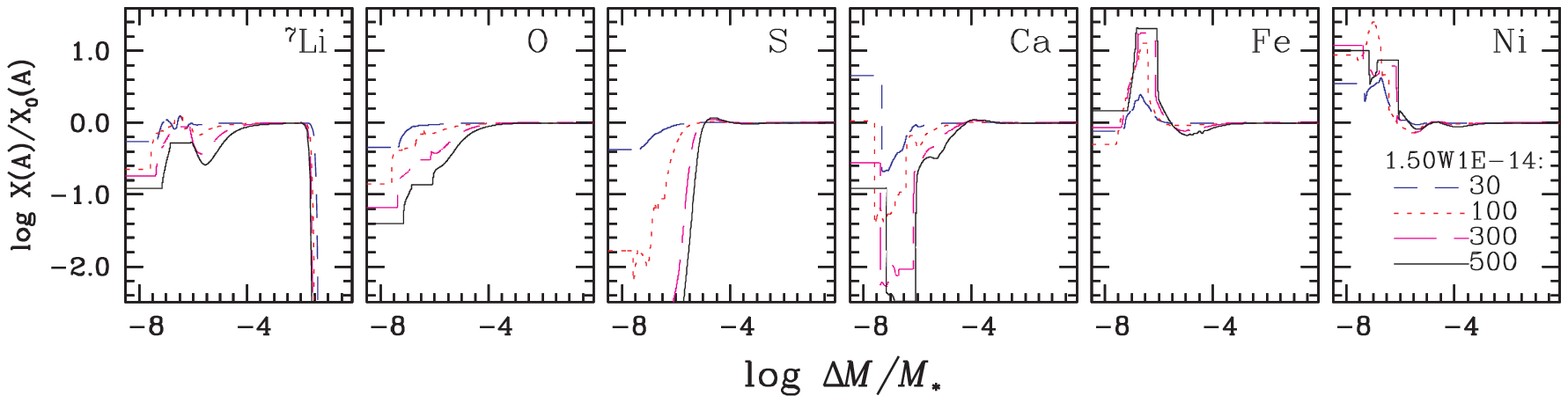}
\caption{[Top row] Radiative accelerations (solid line) and local gravity (dotted line) for a 
few selected elements in a 1.5\,\Msol{} model
with two different mass loss rates. 
For the 1.50W1E-13 model, the radiative accelerations are shown for 
2 ages (30 and 503\,Myr), while they are only shown at 500\,Myr for the 1.50W1E-14 model.
The vertical lines show the position of the bottom of the surface 
H-He convection zone. The corresponding 
internal abundance variations at different ages (in Myr) are shown for both 
the 1.50W1E-13 model [middle row] and the
1.50W1E-14 model [bottom row].
}\label{fig:abint1.5}
\end{center}
\end{figure*}

In Figs.
\ref{fig:abint1.5} and \ref{fig:abint2.5} the radiative 
accelerations as well as the corresponding spatial abundance variations 
for a few selected elements 
are shown for models 
of 1.5\,\Msol{} and 2.5\,\Msol. These
masses approximately correspond to the lower and higher \teff\, limits of AmFm stars. 
 
The MS lifetime of the 1.5\,\Msol{} model is 
about 2\,Gyr whereas the 2.5\,\Msol{} star has a MS lifetime of about 520\,Myr 
(see Fig.\,\ref{fig:hist}). The 1.5\,\Msol{} model with a
mass loss rate of $1\times10^{-14}\Mloss{}$ (see bottom panel of 
Fig.\,\ref{fig:abint1.5}) was stopped at 575\,Myr 
because the solution became numerically unstable. 

Throughout most of the stellar envelope, the mass loss rate does 
not significantly affect the resulting \gr s, as seen in the upper row of Fig.\,\ref{fig:abint1.5}. 
However, if
mass loss is small enough to permit iron peak accumulation below the SCZ, as is the case for the 
1.50W1E-14 model,
then all \gr s will be affected by competition in the region where metals have accumulated.

We present in Sect.\,5.1.1 and Sect.\,5.1.2 approximate equations which will facilitate
the interpretation of the detailed solutions 
shown in Figs. \ref{fig:abint1.5} and \ref{fig:abint2.5}, which will be discussed
in Sect.\,5.2.

\subsubsection{The interior wind solution}
\label{sec:interiorwind}
Consider the approximate solution to Eq. [\ref{eqn:charb2}], in a regime for which the 
$\frac{\partial c}{\partial t}$ term is small compared to the others 
(which is true over most of the MS lifetime). 
Then one may write:
\begin{equation}
\nabla\cdot\mbox{[}-\rho D{\bf\nabla}\ln c+\rho({\bf U}+{\bf U}_{w})c\mbox{]}=0
\label{eqn:int1}
\end{equation}
if nuclear terms are negligeable, which is true for the stellar envelope, and 
$S_{\rm w}=0$, which is true
outside the surface convection zone. 
In one dimension, and in the absence of turbulence (so that $D$ becomes $D_{12}$, 
the diffusion coefficient in a hydrogen and helium background) Eq. [\ref{eqn:int1}]
becomes:
\begin{equation}
\mathcal{F}(r)=r^{2}\left[-\rho D_{12} \frac{\partial c}{\partial r}+\rho(U+U_{{\rm w}})c\right]=cst
\label{eqn:cst}
\end{equation}
where $\mathcal{F}(r)$ is constant as a function of 
$r$ and $U$ is approximately given by Eq. [9] from \citealt{michaud83}, 
with a slight correction\footnote{$\frac{m_{\rm p}}{kT}$ had been erronously forgotten in 
the 2nd and 3rd terms.}:
\begin{eqnarray}
\nonumber U\simeq v_{ \rm D}=D_{12}\left[ -\left(A-\frac{Z}{2}-\frac{1}{2}\right)\frac{m_{\rm
p}g}{kT}\right.+
\\ 
\left.\frac{m_{\rm p}A\gr}{kT} +\alpha _T \frac{\partial \ln T}{\partial r}\right].
\label{eqn:diffusion}
\end{eqnarray}
Here $A$ and $Z$ are the atomic 
mass and charge number 
respectively, 
and $\alpha _T$ is the thermal diffusion coefficient\footnote{This 
equation is not used for the calculations. For more details see \citet{turcotte98soleil}.}.
Since the $g$ and $\gr\,$ terms are multiplied by $A$ 
while the abundance gradient term (in Eq. [\ref{eqn:cst}]) is not, only 
very large abundance gradients can affect the diffusion velocity.
Therefore, the internal abundance distribution approximately satisfies the relation:
\begin{equation}
\mathcal{F}(r)=cst\simeq r^2\rho(U+U_{{\rm w}})c
\label{eqn:cst1}
\end{equation}
which expresses the conservation of particle flux throughout the stellar envelope.
In this discussion, for the purpose of illustration, one may also neglect 
the term involving $\frac{\partial \ln T}{\partial r}$ because it is
not dominant in the relevant regions of the envelope 
(although it is always included in our calculations). 
 
\begin{figure*}[!t]
\begin{center}
\includegraphics[scale=1.01]{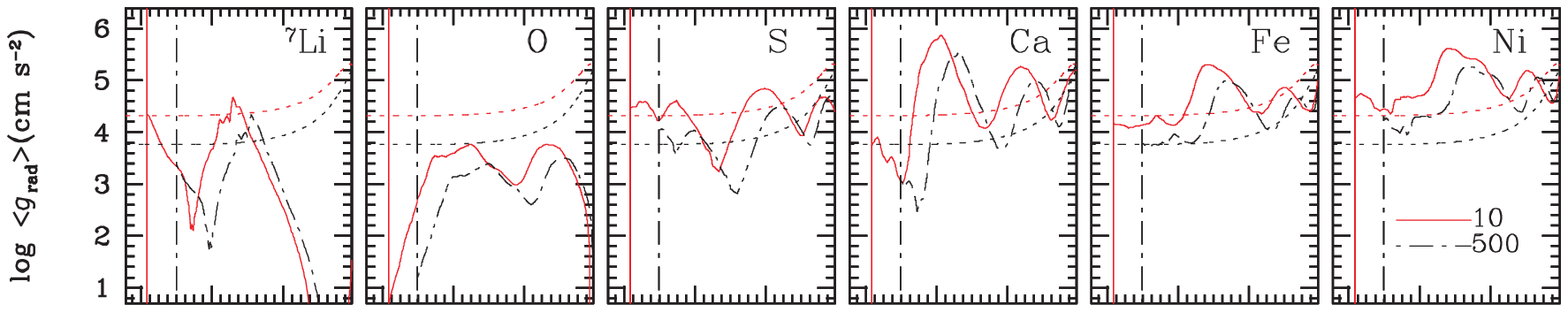}
\includegraphics[scale=1.01]{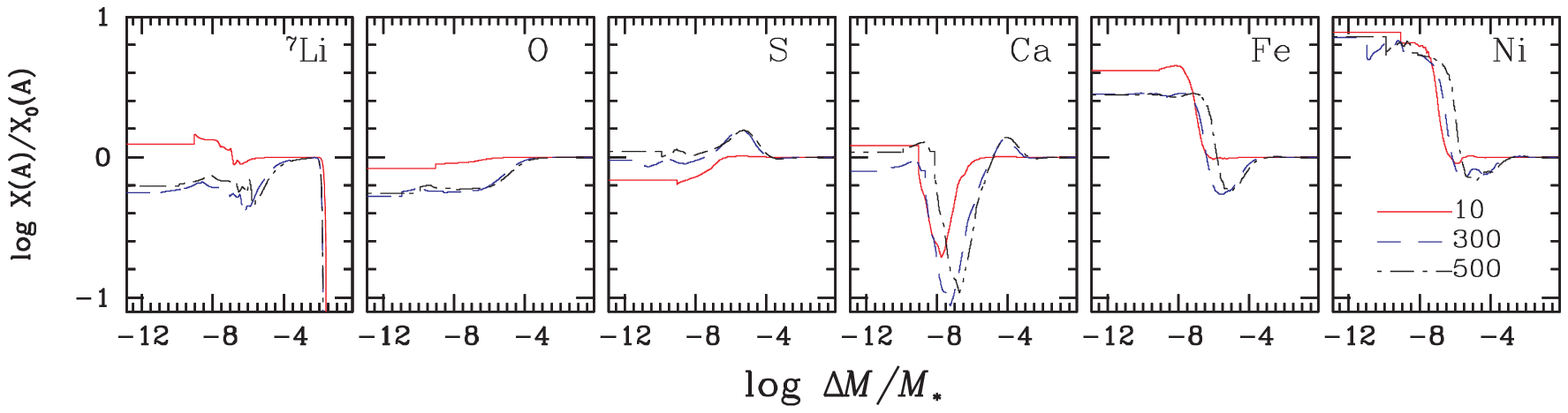}
\caption{[Top row] Radiative accelerations (solid line) and local 
gravity (dotted line) in a 2.5\,\Msol{}
model with a mass loss rate of 
1$\times 10^{-13}$\Mloss. The corresponding internal 
abundance variations are 
shown at three ages [bottom panel].
All curves are identified by their age (in Myr). The vertical lines show the 
position of the bottom of the surface H convection zone.}\label{fig:abint2.5}
\end{center}
\end{figure*}
\begin{figure*}[!t]
\begin{center}
\includegraphics[scale=.80]{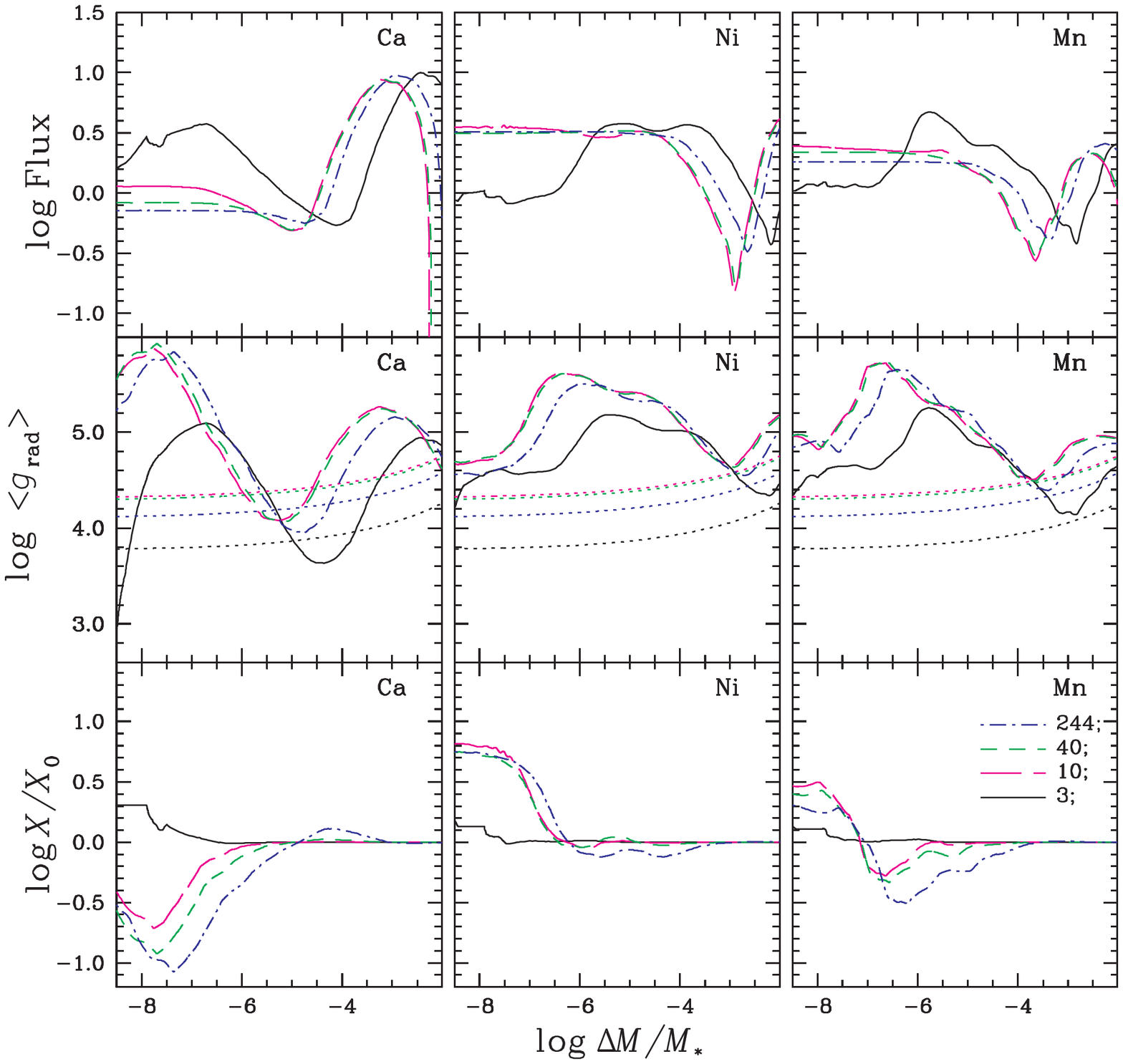}
\caption{Comparison of the normalized local flux with radiative 
accelerations and internal abundances 
for 3 elements at 4 different ages (in Myr) for a 2.5\,\Msol{} model 
with a mass loss rate of 
$10^{-13}$\Mloss.}\label{fig:flux}
\end{center}
\end{figure*}

The implications of flux conservation are illustrated 
in Fig.\,\ref{fig:flux} for three atomic 
species (Ca, Mn and Ni) from the 2.5\,\msol{} 
model of Fig.\,\ref{fig:abint2.5}. 
For all three species, the flux is always positive 
as may be seen from the top row.  While \gr(\Ca) is 
smaller than gravity at some mass shells (middle row), the 
downward diffusion velocity of Ca is never larger than the wind 
velocity (see Fig.\,\ref{fig:vwind}) so that it is dragged by the 
wind toward the surface. In general, as long as the absolute value of the 
diffusion velocity is smaller than the  
wind velocity, 
the concentration simply increases in order to conserve
the flux (as implied by Eq.\,[\ref{eqn:cst1}])\footnote{This resembles the results 
for oxygen which are discussed in \citet{landstreet98}.}. This equation may then be 
used here for all three species.  
As may 
be seen from the bottom row, after 3\,Myr, the concentration 
has adjusted only down to \Dm $\simeq 3 \times  10^{-7}$.
Therefore, below this depth, the local
flux (top row) mainly reflects the local \gr{}. 
After about 10\,Myr however, the abundance has 
adjusted to carry the flux 
which arrives from deep inside the star 
(down to \DM $\simeq -6$). This may be 
evaluated using $\Delta M \simeq \dot M \times t$ for a mass 
loss rate of $10^{-13}$\Mloss.  After 100\,Myr, 
the concentration has adjusted to the flux down to \DM $\simeq -5$ and 
the concentration becomes the mirror image of \gr{} from the surface 
to that depth as may be seen by comparing the middle and bottom rows.
There are Ca
overabundances at $\DM=$ $-8.5$ and $-4.5$, which occur at \gr(Ca)
minima 
(see also Fig.\,\ref{fig:abint2.5}), while the 
minimum abundance occurs at the maximum of \gr(\Ca). 
However, if the wind velocity were not larger than the settling 
velocity, a gradient could develop to conserve the flux. 
If this gradient cannot become large enough to conserve flux, then the 
$\partial c/\partial t$ term (in Eq.\,[\ref{eqn:charb2}]), 
which is required to be
zero in the kinematic approximation
becomes important, and the approximation 
leading to Eq.\,[\ref{eqn:cst1}] becomes invalid.


The net elemental flux toward the 
surface quickly becomes nearly constant 
in time once the local abundance has adjusted itself to conserve 
the flux. The remaining slow variation in time of the flux comes from the
variation of \gr{} where the matter originated, 
deep in the star, at $\Delta M \simeq \dot M \times t$.
  
\subsubsection{The kinematic approximation}

By following the analysis of \citet{michaud86}, 
the surface abundance of elements which are pushed 
upwards by \gr{} and/or dragged by the wind throughout
the stellar envelope can be 
approximated by using a simple kinematic equation {\it so long as the evolutionary 
effects and the contribution of $\frac{\partial c}{\partial t}$ are 
small (i.e. there is no significant accumulation)}. 
The problem of determining surface abundances thereby reduces to a  
kinematic problem. The only required quantity is the local 
velocity and, consequently, the radiative acceleration.  
Elements which originated at $r_1$ at $t=0$ arrive at $r_{cz}$, the 
radius at the bottom of the 
surface convection zone at a time $t_1$ given by:
\begin{equation}
t_1 = \int_{r_1}^{r_{cz}} \frac{dr}{v_{\rm w}+v_{\rm D}}.
\label{eqn:transporttime}
\end{equation}
Because of flux conservation, the flux of an element $A$ entering the 
convection zone at $t_1$ is then given by:
\begin{equation}
\phi (t_1,A)=c_0(A,r_{1})(v_{{\rm D},r_1}+v_{{\rm w,}r_1})\rho_{r_1}\frac{r_1^2}{r_{cz}^2}, 
\label{eqn:fluxCZ}
\end{equation}
where $c_0(A,r_1)$ is the initial abundance 
of $A$ at $r_1$ and $v_{{\rm D}1},\,v_{{\rm w}1}$ and $\rho_1$ are evaluated at $r_1$.
This is valid for the region above which
$v_{\rm w}+v_{\rm D}> 0$ and for $t < t_{0A}$, 
the time required for the element $A$ to migrate 
from the point where $v_{\rm w}+v_{\rm D}\simeq 0$ 
to the bottom of the convection zone.
Since diffusion timescales are much longer 
than convective mixing time scales, the convection zone is assumed thoroughly mixed
up to the bottom of the wind forming region. Within the context of unseparated mass loss, 
the evolution 
of the abundance of $A$ in the surface 
convection zone, $c_{\rm {cz}}(A)$ can then be approximated by: 
\begin{equation}
M_{\rm {cz}}\frac{\partial c_{\rm {cz}}(A)}{\partial t}=- \dot M c_{\rm {cz}}(A)+4\pi r_{\rm {cz}}^2\phi(t_1,A),
\label{eqn:int2}
\end{equation}
where $M_{cz}$ and $r_{cz}$ are 
respectively the mass of the superficial convective zone and the radius at the bottom of this 
convection zone. Eq. [\ref{eqn:int2}] is \emph{not} used for our calculations, 
but is used to interpret the results.

\subsection{Discussion: Internal variations}
 
By comparing Figs. \ref{fig:abint1.5} and \ref{fig:abint2.5}, it is clear 
that the same mass 
loss rate leads to quite different  internal concentration variations
in stars of different mass. There 
are two main reasons for this.
First, in the more massive star, the surface convection zone is much thinner, therefore 
the radiative zone is extended upward into regions where 
the diffusion timescales are much shorter. Second, because of the $\teff ^4$ dependence 
of the photon flux, most \gr s are stronger 
in more massive stars. 

For both models, diffusion mainly affects the outer 
$10^{-3}$ of the star's mass; the point above which 
the effects of atomic diffusion become visible on 
Figs. \ref{fig:abint1.5} and \ref{fig:abint2.5} will be defined as 
\emph{the point of separation}.  Atomic diffusion can also act 
deeper in the star (e.g. in the core), though in A and F \MS{} stars,
its effects are much smaller below than above the point of separation.

\subsubsection{The 1.5\,\Msol{} models}
\label{sec:abint1.5}
For the 1.5\,\Msol{} model with a mass loss rate of $1 \times 10^{-13} \Mloss\,$ 
(henceforth designated 1.50W1E-13 for short), internal anomalies 
are small. 
Iron reaches an overabundance of 1.5, and nickel of 2.0. 
Lithium, oxygen and calcium 
decrease from the point of separation up to the bottom of the 
surface convection zone. Differences in \gr{} 
lead to differences in their behavior. Due to a strong \gr{} immediately below the surface CZ, 
Ca is most underabundant in that region, 
since Ca is pushed into the convection zone by both \gr{} 
and the wind while little arrives from below because of the minimum in its \gr{} 
at $\DM \sim -4.5$. At 500\,Myr, 
its underabundance reaches a factor of 4 while the Li 
underabundance below the SCZ is about $-0.3$\,dex. 

\begin{figure*}[!t]
\begin{center}
\includegraphics[scale=.80]{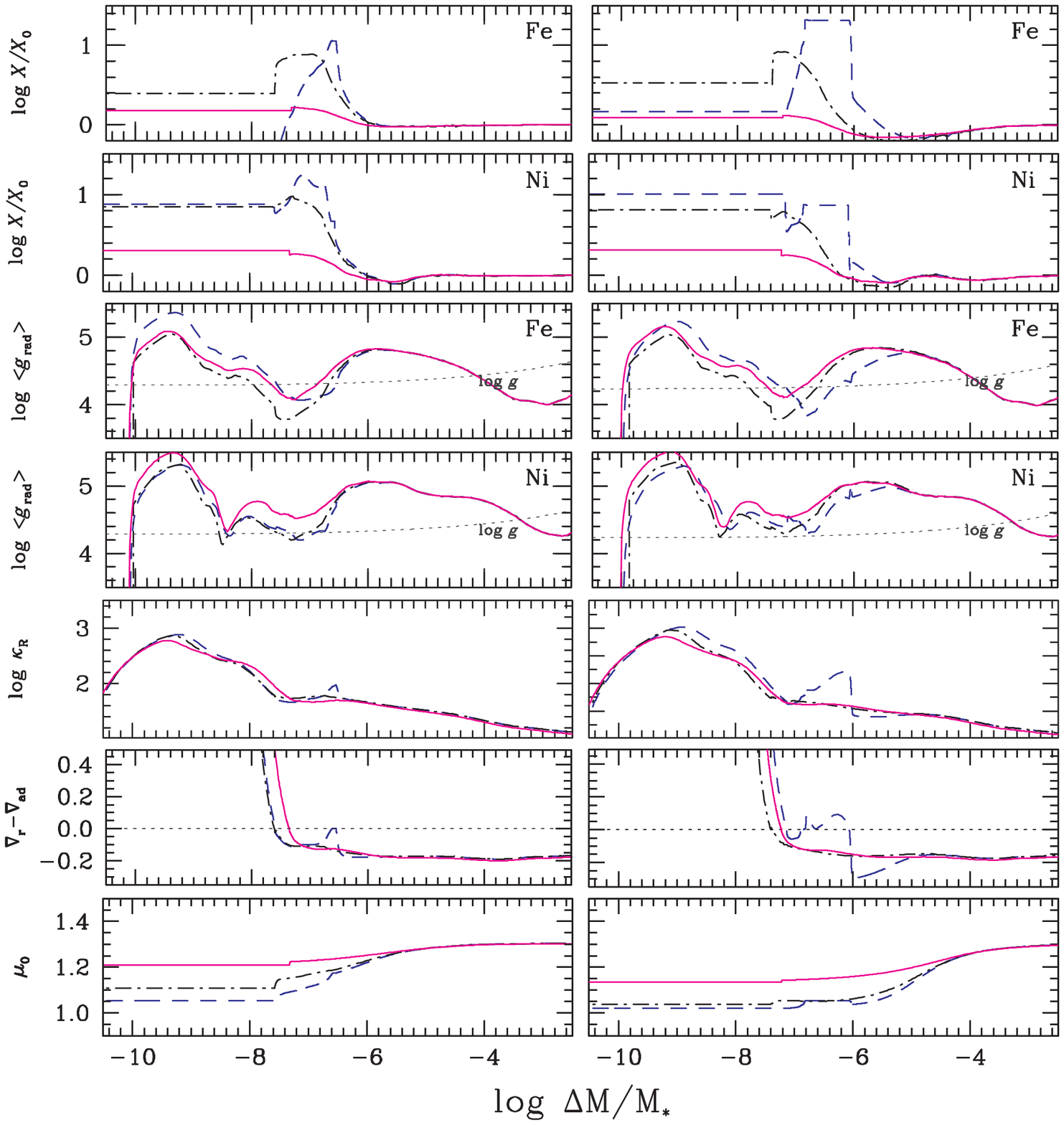}
\caption{ Internal abundances (Fe, Ni), radiative accelerations (Fe, Ni), 
Rosseland opacity, the difference between the radiative and adiabatic 
temperature gradients as well as the mean molecular weight per nucleus for three 
1.5\,\Msol{} models with different mass loss rates ($10^{-14}\Mloss$ [dashed line],
$2 \times 10^{-14}\Mloss$ [dot-dashed line] and $10^{-13}\Mloss$ [solid line]) at 70\,Myr
(left column) and 500\,Myr (right column). 
At this mass, the Fe peak convection zone (between $\DM = -6$ and $-7$) 
only appears for the mass loss rate of $10^{-14} \Mloss$. 
The H-He convection zones remain linked for all mass loss rates.}\label{fig:kabmu1.5}
\end{center}
\end{figure*}

In the lower row of Fig.\,\ref{fig:abint1.5}, one sees 
that when the wind is 10 times smaller, the internal variations are much 
stronger since the 
advection by $v_{\rm{wind}}$ is not strong enough to prevent 
elemental accumulation. Lithium has an interesting behavior: it has 
a local maximum where its \gr{} equals gravity.
The underabundances of Li and O drop to $-0.95$ and $-1.35$ dex respectively, 
while underabundances
greater than 2\,dex are reached for S and Ca. 
Through Eq. [\ref{eqn:cst1}], 
when $U_{\rm w}$ is smaller, the change in $U$ due to \gr{} has a larger effect on concentration. This 
is particularly 
true for Fe: due to a dip in \gr{} (see Fig.\,\ref{fig:abint1.5}), it accumulates 
near $\DM=-6.5$ (where $T=200\,000\,$K). The 
overabundances of Fe and Ni reach 1.25
and 1.45\,dex respectively. The implications of this accumulation 
are analyzed in Fig.\,\ref{fig:kabmu1.5}.
The local opacity bump in the region between $\DM \simeq 6.0-6.8$ leads to 
the appearance of an 
iron peak convection zone ($\nabla_r-\nabla_{ad}$ changes sign) at
around 70\,Myr  
which survives until the end of the simulation. Fe and Ni reinforce each other.
Only the $10^{-14} \Mloss\,$ model becomes convective; doubling the mass 
loss rate reduces the Fe accumulation enough
for the iron peak convection zone not to appear. 

In fact, a noteworthy transition occurs in 
internal solution
types between the 1.50W1E-14 and 1.50W2E-14 models (Fig.\,\ref{fig:kabmu1.5}). In the 
1.50W1E-14 model, Fe is overabundant where \gr{(Fe)}\,$>g$, 
whereas Fe is minimal at the \gr{(Fe)} maximum in 
the 1.50W2E-14 model (similarly, Ca is minimal at the \gr{(Ca)} maximum in
the 2.50W1E-13 model shown in Fig.\,\ref{fig:flux}). This is due to the relationship between
$v_{{\rm wind}}$ and the settling velocities for each 
individual species; if the settling velocity
dominates locally, the element will accumulate locally, and the 
solution will behave as does Fe 
in the 1.50W1E-14 model. When $v_{\rm wind}$ dominates throughout the envelope, 
the solution is determined by flux conservation, as shown in Fig.\,\ref{fig:flux}.
This transition 
separates the solution in which the surface abundances reflect 
matter which is advected from deep inside the star ($v_{\rm wind}$\,$>$\,$v_{\rm sett}$) from 
the solution in which the surface reflects variations at the bottom of the SCZ (as in the models of
\citealt{watson71} and \citealt{alecian96}). This transition occurs only for elements
whose \gr{} has a minimum smaller than gravity close to the bottom of the SCZ. 
 
A large enough mass loss rate also 
affects the position of the bottom of the surface H-He convection zone 
by keeping He 
from sinking. Since it also modifies the accumulation of metals, mainly iron and nickel,
its effect is complex:  the relative position of the bottom of the convection zone for the 
three mass loss rates in Fig.\,\ref{fig:kabmu1.5} is different at 70 and 500\,Myr. 

The inversion of the molecular
weight gradient (bottom row of Fig.\,\ref{fig:kabmu1.5} at 500\,Myr), which eventually follows the 
appearance of the iron accumulation around 
200\,000\,K, has been suggested to affect the presence
of the iron peak convection zone (\citealt{theado09}). 
In the present calculations however, the iron peak convection
zone appears (as seen at 70\,Myr) before the inversion
appears so that while the size of the convection 
zone could be affected by the $\mu$ gradient inversion, its 
appearance cannot be affected. This will be further discussed in Sect.\,\ref{sec:conclusions}.

\subsubsection{The 2.5\,\Msol{} models}
\begin{figure*}[!t]
\begin{center}
\includegraphics[scale=.80]{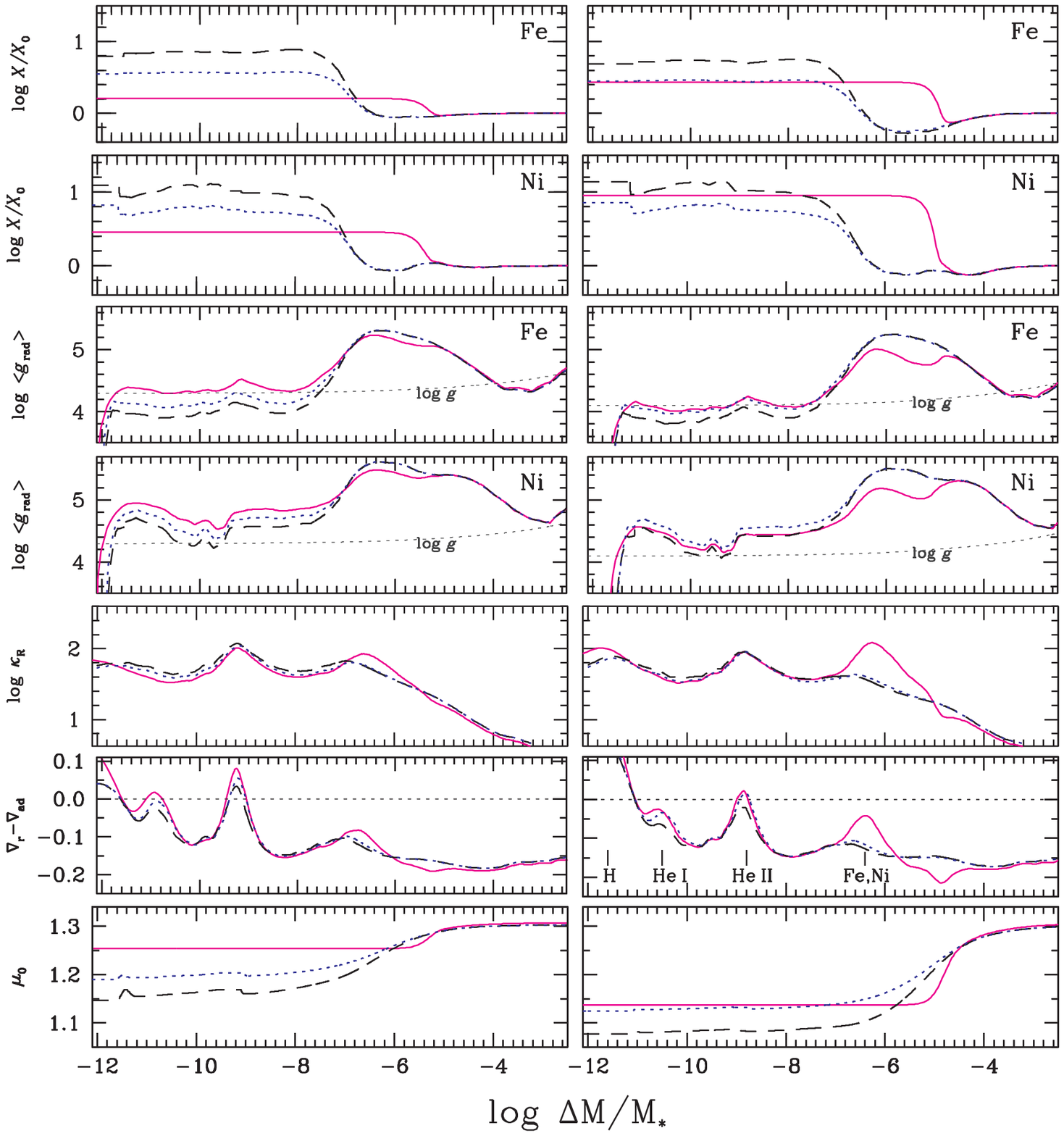}
\caption{Internal abundances (Fe, Ni), radiative accelerations (Fe, Ni), 
Rosseland opacity, the difference between the radiative and adiabatic 
temperature gradients as well as the mean molecular weight per nucleus for two 
2.5\,\Msol{} models with different mass loss rates ($10^{-13}\Mloss$ [dotted line] 
and $5 \times 10^{-14}\Mloss$ [dashed line]) as well as a model with turbulence [solid line]
at 30\,Myr (left column) and around 250\,Myr (right column). The maxima in the 
$\nabla_{\rm r}$-$\nabla_{\rm ad}$ row are mainly due to opacities from 
H, He\,I, He\,II and Fe/Ni from left to right
respectively.}\label{fig:kabmu2.5}
\end{center}
\end{figure*}
Large overabundances of iron peak elements are obtained in 
the 2.5\,\Msol{} models (Fig.\,\ref{fig:abint2.5}): Ni and Fe are 
overabundant by factors of 7.5 and 4 respectively. Calcium 
on the other hand is found to be up to 10 times underabundant around \DM=$-6.5$, 
where its \gr{} is near a maximum. This may seem counter intuitive but  is a consequence 
of Eq.\,[\ref{eqn:cst1}]. Since the flux is conserved, where 
$v_{\rm D}$ is large and positive, $c$ decreases,
which is precisely what happens for Ca just
below the convection zone (see Figs. \ref{fig:abint2.5} 
and \ref{fig:flux} and the last two paragraphs of Sect.\,\ref{sec:interiorwind}).
\begin{figure*}[!t]
\begin{center}
\includegraphics[scale=.98]{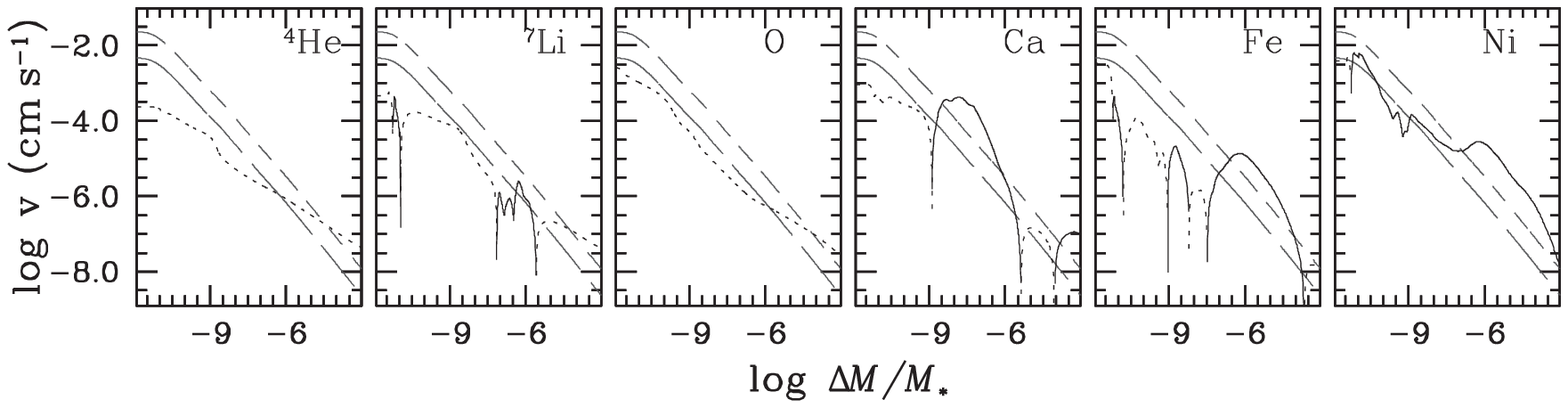}
\caption{Wind velocities (long dashed line: $10^{-14}$\Mloss;
dashed line: $10^{-13}$\Mloss) and diffusion 
velocities (solid when positive, towards the surface, and dotted when 
negative) 
of a few selected elements in two 2.50\,\Msol{} models at 300\,Myr. For most species, wind
velocities decrease more rapidly inwards than diffusion velocities.}\label{fig:vwind}
\end{center}
\end{figure*}



Wind and diffusion velocities  are compared in 
Fig.\,\ref{fig:vwind}. The wind velocity decreases more 
rapidly than the diffusion velocity as \DM{} increases.  For instance, 
the absolute value of the $^4\He$ settling velocity is 50 times 
smaller than the larger of the two wind velocities at $\DM = -12$ but equals it at $\DM = -4.5$.  
For smaller mass loss rates the settling velocity dominates closer to the surface.  
Calcium has an upward diffusion velocity over the interval $-9 < \DM < -6$ and it 
is up to ten times larger than the wind velocity. Fe and Li have smaller 
upward diffusion. Those velocities determine local concentration via Eq.\,[\ref{eqn:cst1}]
(see the end of Sect.\,\ref{sec:interiorwind}).

Fig.\,\ref{fig:kabmu2.5} allows an analysis of the effect of varying 
mass loss rates on surface convection zones
and to distinguish the effects of mass loss from those of turbulence 
(see Sect.\,\ref{sec:turb}). The abundances and \gr{} of both Fe and Ni are 
presented since they are the main contributors to the appearance of the iron 
peak convection zones in 1.5\,\Msol{} models. In 2.5\,\Msol{} models however, iron convection 
zones do not appear for the two mass loss rates considered.  The Fe and Ni abundances
remain slightly below solar where \gr{(\Fe)} and \gr{(\Ni)} are largest so 
that they do not lead to an increase of opacity where they could contribute most 
to opacity. The wind velocity is ten times larger than the diffusion velocity of \Fe{} 
for $\DM \ge -8$ so that Eq.\,[\ref{eqn:cst1}] forces the solution to be nearly 
constant over that range.
This is to be contrasted with their behavior in 1.5\,\Msol{} models.  

Even a small mass loss rate difference can have important 
effects on the stellar structure in  2.5\,\Msol{}
models. 
For both  mass loss rates considered, the He\,I convection zone disappears 
early in the evolution, at
around 30\,Myr (see the fifth row of Fig.\,\ref{fig:kabmu2.5}). 
The He\,II convection zone on the other hand only 
disappears in the  model with the smaller mass loss rate, the 2.50W5E-14 model, at around 200\,Myr. 
This is because the 
inward diffusion velocity of He 
dominates the wind velocity at a shallower
depth than in the 2.50W1E-13 model 
(see Fig.\,\ref{fig:vwind}),
so that a more pronounced He underabundance develops to conserve the flux.


\subsection{Surface abundance variations}
\label{sec:surfab}
\subsubsection{The 1.5\,\Msol{} models}
\label{sec:surfab1.5}
\begin{figure*}[!t]
\begin{center}
\includegraphics[scale=1.04]{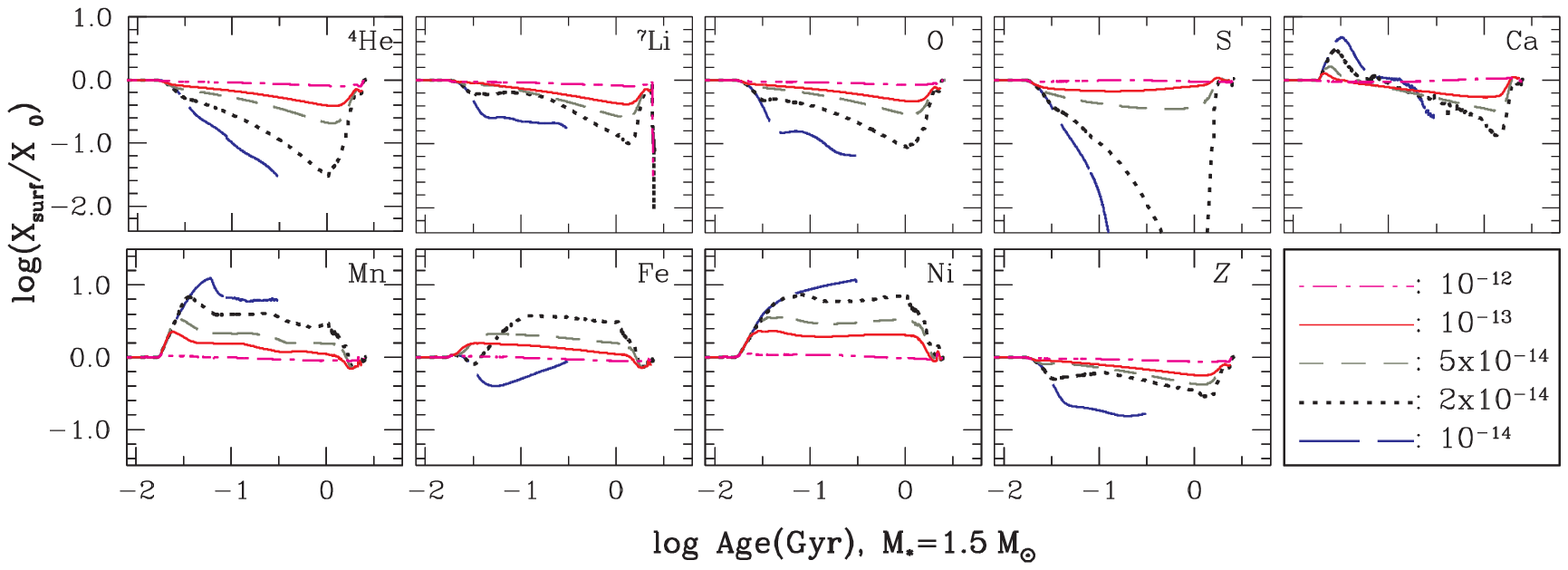}
\includegraphics[scale=1.04]{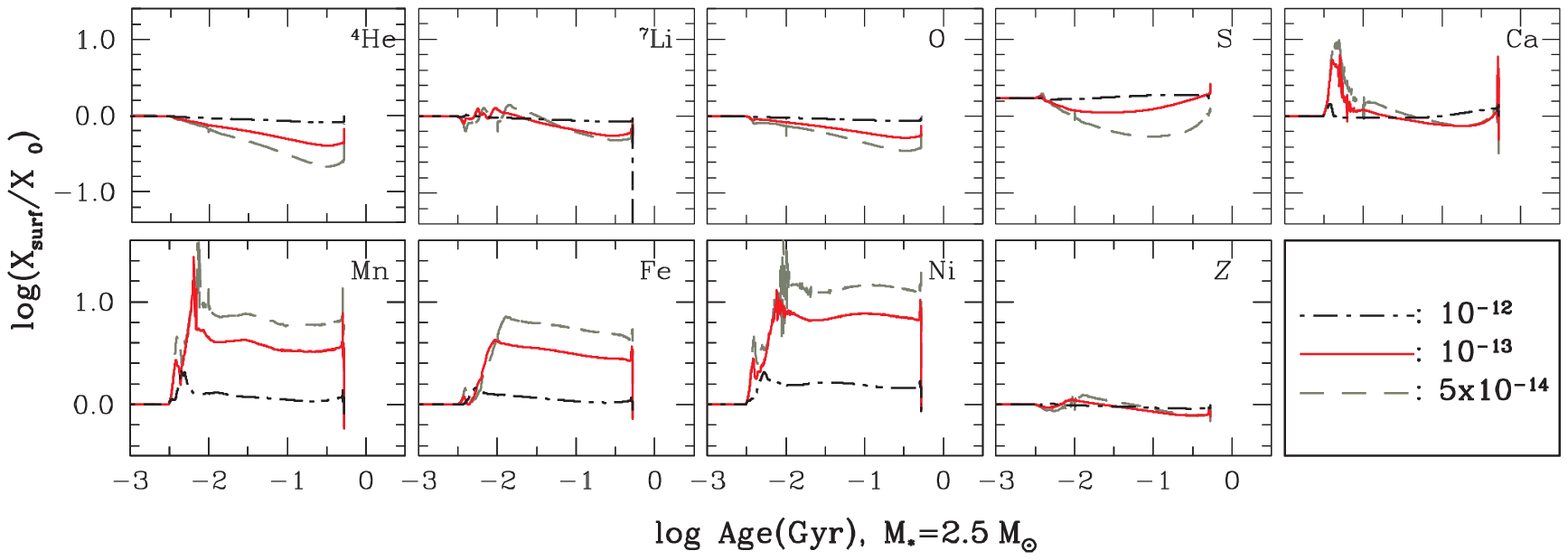}
\caption{Surface abundance variations for five 1.50\,\Msol{}
models [top 2 rows] as well as three 2.50\,\Msol{} [lower 2 rows] 
models with $Z_0=0.02$ and varying amounts of mass loss (in \Mloss.). 
The  model was ended at 
around 550\,Myr due to numerical 
instabilities which were caused by the severe 
underabundances of many elements
(of which S is shown). It is also the only model 
which leads to the appearance of a separate iron 
peak convection zone. In the bottom panel, 
the two models with smaller mass loss rates 
lead to a surface metallicity which 
has a maximum around 10\,Myr 
before falling below $Z_0$ around 100\,Myr.}\label{fig:absurf}
\end{center}
\end{figure*}

The evolution of surface abundances is 
shown for several 1.5\,\Msol{} models in 
the upper two rows of Fig.\,\ref{fig:absurf}. For the 
1.50W1E-12 model, the largest anomaly 
encountered, which is for Ni, reaches a mere 
0.05\,dex, and it is quickly flattened after 
a brief period. Such a mass loss rate 
effectively wipes out any surface effects of chemical separation. 
In the 1.5W1E-14
model, some of the calculated underabundances are very large 
(greater than 2 dex for S as well as C and Si, which are not shown).
With the exception of Fe, all iron peak elements are 
overabundant throughout the simulation. 
Both Ni and Mn become more than 1\,dex overabundant 
at around 500\,Myr and 50\,Myr respectively. 

Iron surface
abundances are particularly interesting for this model since it is the 
only one to have an iron surface underabundance 
which spans 500\,Myr\footnote{We 
may assume however that if the simulation had not encountered 
significant
instabilities, the shown iron abundance 
trend would have continued and iron would have become overabundant.}.
This is due to the accumulation of iron under the surface convection zone, and the 
subsequent appearance of an
iron peak convection zone (see Fig.\,\ref{fig:kabmu1.5}). Although not shown here, a separate 
iron peak
convection zone is expected for all simulations for masses between 1.47\,\Msol{} and
3.0\,\Msol{} when the mass loss rate is equal to or below 
$\sim 10^{-14}\Mloss$. 

\begin{figure}[!h]
\begin{center}
\includegraphics[scale=.46]{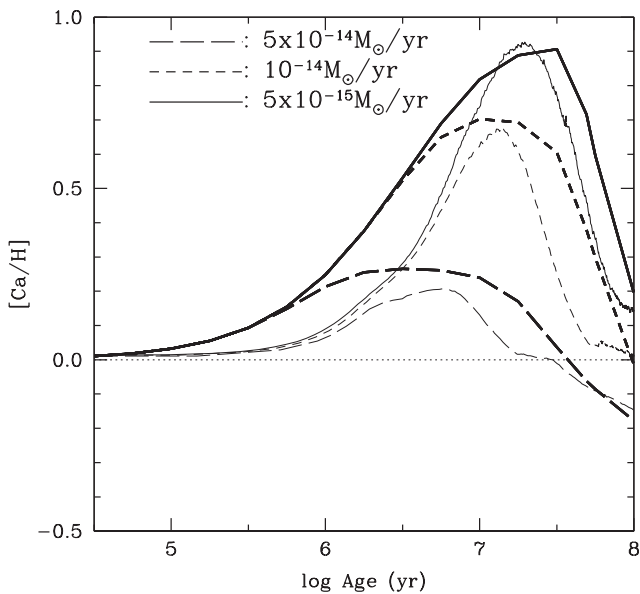}
\caption{Evolution of Ca surface abundances for 1.5\,\Msol{} models 
with different mass loss rates. The calculations from \citet{alecian96} are shown in bold.
}\label{fig:alecian}
\end{center}
\end{figure}

The 1.50W2E-14 and 1.50W5E-14 models generate surface iron 
overabundances of about 0.5 and 0.35 respectively at 300\,Myr. 
The calcium abundance is particularly 
interesting since for all models, an overabundance is predicted at the 
beginning of the main sequence 
evolution followed by an underabundance. The underabundance 
is present over a much larger fraction of the evolution than the overabundance.   The Ca calculations are compared in  Fig.\,\ref{fig:alecian} with those
carried out by \citet{alecian96}. His calculations were undertaken
in static stellar models; therefore, evolutionary 
effects were not included. There are also slight difference in \teff\, between our calculations 
($\teff \simeq 7200$\,K at the beginning of diffusion but 
decreases as the star evolves) and his simulations 
($\teff=7500$\,K). There are also differences in \gr(Ca): 
his calculated \gr{(Ca)} has a 
peak which is 10 times smaller than obtained in a 
model of similar \teff{} with our code. Results are compared 
in Fig.\,\ref{fig:alecian} for three mass loss rates.  In all three
cases, the maximum anomaly as well as 
the overall behavior of the curves correspond well. The overabundance 
peaks are however not quite so wide in our calculations as in his. The agreement seems satisfactory.


\subsubsection{The 2.5\,\Msol{} models}
\label{sec:absurf2.5}
In the 2.5\,\Msol{} models (lower two rows of Fig.\,\ref{fig:absurf}), 
the surface abundances of heavier metals (say $Z > 17$) 
can be characterized by two distinct episodes: a steep abundance spike, 
followed by a smooth decline.
The initial peaked episode around 10\,Myr shows matter which was above 
$\DM \simeq -6$ at the onset of diffusion,  advected to the surface 
 by the stellar wind. 

As  discussed at the end of Sect.\,\ref{sec:interiorwind}, 
the initial  flux
distribution reflects \gr{} since, originally, 
the concentration is the same throughout the envelope; then, as time elapses, the 
internal abundances naturally evolve in such a way that flux is conserved 
throughout the envelope (see Fig.\,\ref{fig:flux}). In so far as the $\partial c/\partial t$ term is negligible 
(i.e. flux is conserved),
the surface reflects the point in the initial flux distribution which 
is dragged by $v_{\rm{wind}}$. Therefore, when the initial internal flux profile of an element has
important variations near the surface due to strong variations 
in \gr{} (as is the case for Ca and Ni, 
see Fig.\,\ref{fig:abint2.5} and \ref{fig:flux}), these variations  appear
on the surface in a time which is related to the mass loss rate.  

Quantatively, in the
2.50W1E-13 model (solid line), a given element's surface abundance at 100\,Myr 
depends on the initial flux
variations (caused by \gr{} variations at $t=0$)\footnote{Remember 
that at $t=0$, the flux distribution is essentially
proportional to the local abundance multiplied by the local \gr{} since the composition is homogeneous.} at a 
depth of $\Delta M \simeq \dot M \times t= 10^{-13} \Mloss \times 10^{8}\,{\rm yr} \simeq 10^{-5}
\Msol$. In Fig.\,\ref{fig:absurf},  the slight bump in surface Ni 
abundance around 100\,Myr for the 2.50W1E-13 model reflects the 
small bump in the initial flux
distribution of Ni around $\DM \simeq -5$ (see Fig.\,\ref{fig:flux}). 

The previous example is an application of Eq.\,[\ref{eqn:transporttime}] 
and shows that the time it takes for a 
given internal variation to reach the surface is inversely
proportional 
to the mass loss rate.  This is further 
illustrated by the fact that the surface 
Ni abundance peak around 10\,Myr appears earlier as 
the mass loss rate increases (see Fig.\,\ref{fig:absurf}). 
The dilution by the convection zone (Eq.\,[\ref{eqn:int2}])
is relatively small because it has a relatively small mass.
Also, the rapid variations seen around this Ni abundance peak reflect the
many internal flux variations which were above 
\hbox{$\DM \simeq -6$} at $t=0$. For a mass loss rate of  $10^{-13} \Mloss$, these variations  
all reach the surface within $10^{7}$ yr. 

As a consequence, the nearer an internal flux variation is 
to the surface at $t=0$, the quicker it appears at the
surface, and the quicker it  disappears. A flux variation that spatially 
spans from the surface to $\DM=-7$ at $t=0$ appears and disappears at the surface in less
than one million years for a mass loss rate of $10^{-13} \Mloss$.

The amplitude of the surface variations  also 
depends on the mass loss rate;
as the mass loss rate increases, each internal variation of the flux at $t=0$ 
manifests itself at the stellar surface with a smaller 
amplitude since more enriched/depleted 
matter is evacuated from the SCZ (Eq.\,[\ref{eqn:int2}]). This is illustrated by the 
decreasing amplitude of the initial surface abundance
spikes for Mn and Ni as the mass loss rate increases (Fig.\,\ref{fig:absurf}).

The conservation of the flux down to $\DM\,\simeq -5.5$ to $-6.5$ 
(depending on the age and mass loss rate) has another important consequence: 
diffusion between the surface convection zones (i.e. between the H and 
either of the two He convection zones) has practically no influence on the surface abundances. 
Matter which originates from between the H and He convection zones ($\DM \simeq -10$) 
appears at the surface within
$10^3$ years for a mass loss rate of $10^{-13}$\Mloss{}. The detailed calculation of 
chemical transport between these
zones is then not required in order to accurately obtain the 
surface abundance solution. To verify 
the accuracy of this
assertion a model was calculated with homogenized abundances between all 
surface convection zones, and it was found that
the surface solution was practically identical to the surface solution 
obtained when separation was allowed between SCZs. 
After a mere 3\,Myr of a star's \MS{} life 
with a mass loss rate of $10^{-13}\,$\Mloss, anomalies which appear 
at the surface  reflect the
separation which occurs below $\DM={-7}$.
Hence, \emph{for most of the \MS{} lifetime of models 
compatible with observations, 
the surface 
abundance solution depends on the separation 
which takes place around $\DM\,\simeq -5.5$ to $-6.5$}. 

This is analogous to the turbulence models of \citet{richer00}, 
in which surface abundances depend solely on the separation 
which occurs below 200\,000\,K. 
In our calculations, however, abundance variations are present 
throughout the stellar envelope  because no mixing is 
enforced outside of convection zones. In that respect, in the 
presence of mass loss, it is clear that 
the chemical separation responsible for the AmFm 
phenomenon involves up to $\DM = -5$ of the star's mass. 

\subsubsection{The effect of $Z$, age and \teff}
\label{sec:Effets}
\begin{figure*}[!t]
\begin{center}                                                                                
\includegraphics[scale=1.04]{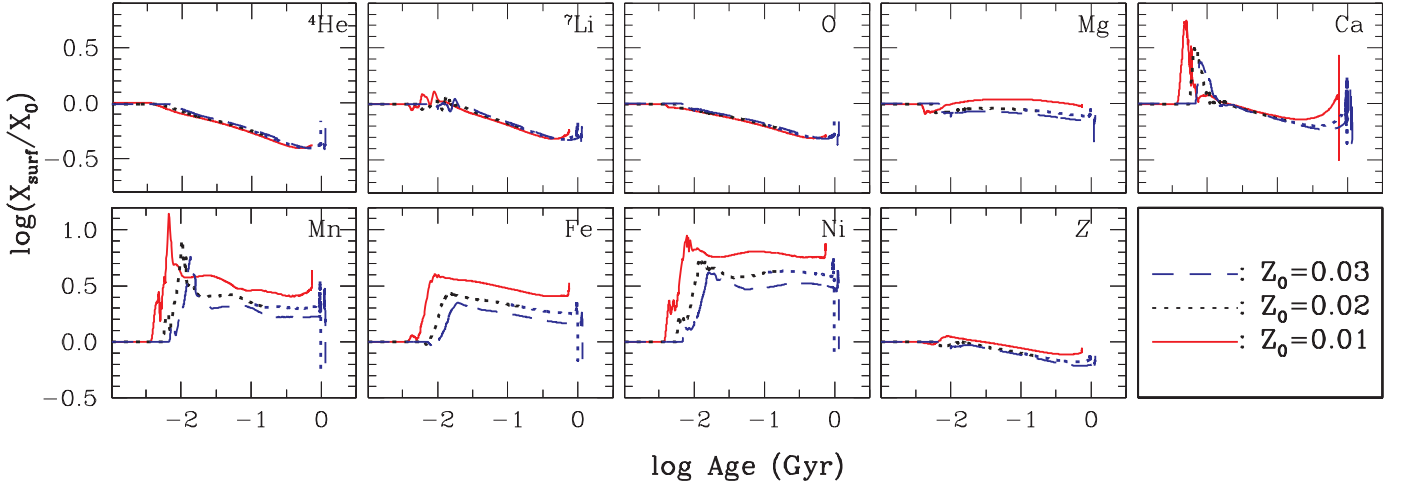}
\caption{The effect of varying initial metallicity 
on the evolution of abundance anomalies at the surface 
of a 2.0\,\Msol{} star with a mass loss rate of $10^{-13} \Mloss$. 
Models are shown for $Z_0=0.01$ (solid), $Z_0=0.02$ (dashed line) 
and $Z_0=0.03$ (long dashed line). The original abundance of Li is assumed 
independent of $Z_0$.}\label{fig:EffetZ}
\end{center}
\end{figure*}

In Fig.\,\ref{fig:EffetZ}, one sees that the main
features of the time evolution of surface 
abundances are similar for three different 
initial metallicities. 
However varying the initial 
metallicity can have an important effect on the amplitude of surface anomalies, 
though not for all
elements. 
For elements such as CNO, 
varying $Z_0$ has 
relatively little effect on the surface abundance anomalies
since these elements are not supported by the radiation field. 
For heavier iron peak elements,  
supported by the radiation field, the 
ramifications are much more apparent since 
their lines are often saturated and flux sharing becomes prevalent. 
In fact, the lower metallicity models have larger
anomalies, when compared to the original abundances of the model, 
since the same flux is shared among fewer atoms. 
Also, the abundance peaks appear earlier during evolution
as $Z$ decreases since the \gr{} profiles with 
respect to $\Delta M/M_*$ are shifted toward the surface as 
$Z$ decreases. However the situation is different if one compares 
to a fixed set of abundances.  
By comparing Figs.\,\ref{fig:absurf} and \ref{fig:EffetZ} one 
notes that a reduction of mass loss by a factor of 1.5 approximately 
compensates for a reduction of $Z_0$ from 0.03 to 0.02 for 
the absolute final abundance of Fe and other elements which are supported. 
It however amplifies the effect of the reduction for elements 
which are not supported by \gr{} such as CNO. 

\begin{figure}[!t]
\begin{center}                                                                                  
\includegraphics[scale=.52]{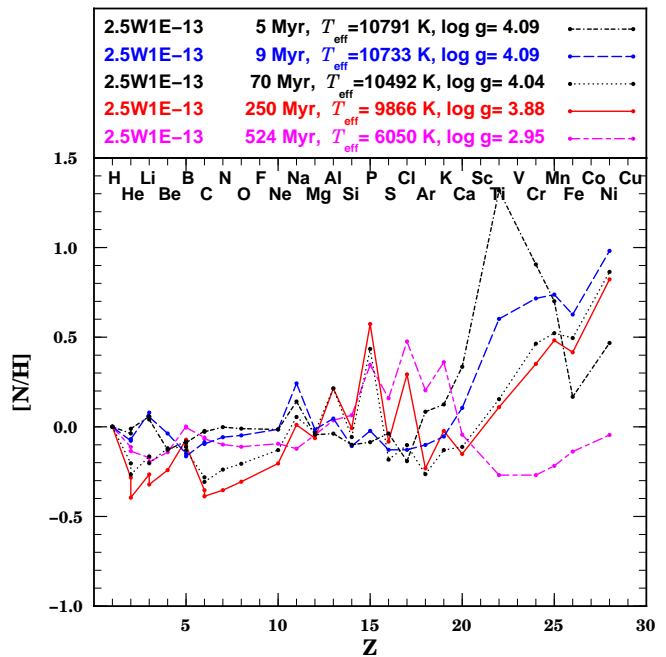}
\caption{Surface abundance anomalies at five ages (5, 9, 70, 250 and 524 Myr) for a 2.50\,\Msol{}
star with a mass loss rate of $10^{-13} \Mloss$.}\label{fig:EffetAge}
\end{center}
\end{figure}

Figure\,\ref{fig:EffetAge} shows the surface abundances as a function of atomic number for  
a 2.5\,\Msol{} model at five
different ages. A large overabundance 
of Ti (1.3\,dex) occurs as early as 5\,Myr because  
\gr(\Ti) has a maximum  near the surface, and 
the bump it causes in the original flux is dragged by the wind. 
The same occurs for Cr and Mn but to a lesser extent. There is 
also a slight overabundance of Li which
appears at this age. At 9\,Myr, the 
Ti overabundance has weakened and 
most iron peak elements reach their highest values: Fe peaks at 
0.65\,dex and Ni peaks at 1.0\,dex.  Throughout the star's \MS{} evolution, the 
surface abundances change only
slightly, as highlighted by the similarity in the curves at 70 and 250\,Myr. 
Until the final dredge-up (i.e. the post turn-off increase in SCZ mass), which begins around 
520\,Myr, the iron peak abundances decrease only slightly. For instance, 
Fe only drops 0.15\,dex over the entire \MS{} lifetime. The changes will 
be slightly more significant
for underabundant elements because they are not supported by the radiation field, and so 
the amplitude of the underabundances is largely dependant on time (due to settling). For CNO, 
the surface underabundances appear around 9\,Myr and continuously 
decrease down to about $-0.40$, $-0.38$ and $-0.35$\,dex for C, N and O respectively, 
after which dredge-up begins. 
\emph{At all times, the surface abundances reflect the matter which 
is exposed by the wind at the specific age}. With that in mind, one 
notes from the curve at 524\,Myr,
that the material at $\DM\simeq -2$ has not yet reached the surface 
since LiBeB abundances have not yet dipped 
due to material advecting to the surface 
from regions where they burn (see Fig.\,\ref{fig:abint2.5}). 
At this age, the surface convection zone extends down to
$\DM = -4.5$, and while 
iron peak elements are now underabundant, an overabundance 
has appeared for S (and a few species of similar atomic mass) because 
of the dredge up of the maximum of its abundance at $\DM \sim -5.2$ 
(see Fig.\,\ref{fig:abint2.5}, one also notes the underabundances of Fe and Ni near 
$\DM \sim -5$ which explain their underabundance at 524\,Myr). This time dependent 
behavior is compared to models with turbulence in Sect.\,\ref{sec:turb}.
\begin{figure}[!h]
\begin{center}                                                                                  
\includegraphics[scale=.52]{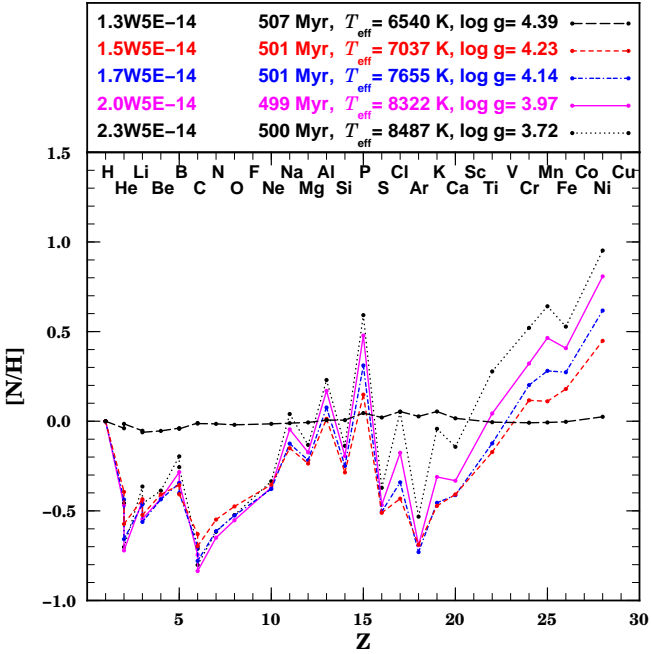}
\caption{The effect of varying \teff{} (or stellar mass) on 
the surface abundance profiles of 5 models at 500\,Myr for a mass loss rate of 
$5 \times 10^{-14}\Mloss$. 
}\label{fig:EffetTeff}
\end{center}
\end{figure}

Figure\,\ref{fig:EffetTeff} shows the superficial abundances at 500\,Myr for 
models of different masses
with the same  mass loss rate (\hbox{$5 \times 10^{-14}\Mloss\,$}). 
The 1.3\,\Msol{} has much smaller 
anomalies than the others. At a given age, stars of 1.5 to 2.3\,\Msol{} have 
very similar underabundances of elements from He to Ne, since for a 
given mass loss rate, they essentially
depend on time. 
Iron peak
overabundances increase as \teff{} increases. 
This is partly caused by the increase of \gr{} 
with \teff{} as well as by the reduction 
of the mass of the surface convection (if the SCZ is more massive, the anomalies 
will be reduced by dilution). This 
behavior is different from what is obtained in turbulence models, and 
will also be discussed in 
Sect.\,\ref{sec:turb}.

\subsubsection{Stars of the lithium gap}
\label{sec:lithium}
\begin{figure}[!b]
\begin{center}
\includegraphics[scale=.42]{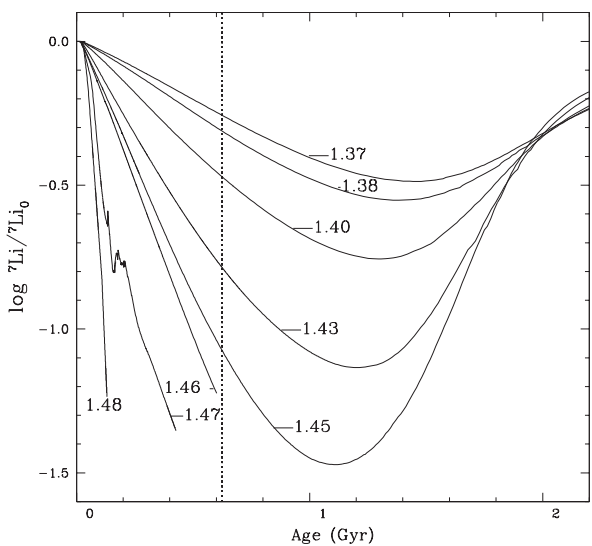}
\caption{Evolution of $^7$Li 
surface abundances for models with masses ranging from 1.37\,\Msol{} to 1.48\,\Msol{} with
no mass loss and $Z_0=0.02$. The vertical line indicates 625\,Myr, the age of the
Hyades open cluster. This can be compared to Fig.\,6 of \citet{richer93}.
}\label{fig:Lisurf}
\end{center}
\end{figure}

The surface Li abundances for stars of the lithium gap without mass loss 
are shown in Fig.\,\ref{fig:Lisurf}. 
The models heavier than 1.45\,\Msol{} ceased converging before the end of their \MS{}
lifetime because of numerical instabilities. 
In the absence of mass loss, the H-He and iron peak 
convection zones are split by
the appearance of a 
thin radiative zone for all models of 1.47\,\Msol{} or more. 
Variations within this layer appear 
at the surface of the 1.47\,\Msol{} model near 200\,Myr.

A close inspection
of Fig.\,\ref{fig:Lisurf} shows a disctinct separation between the 1.46\,\Msol{} 
and 1.47\,\Msol{} models; the 1.46\,\Msol{} model behaves more 
like the 1.45\,\Msol{} model
while the 1.47\,\Msol{} model's behaviour most resembles 
the 1.48\,\Msol{} model. Again, this is
because of the appearance of a radiative zone immediately 
above the iron peak CZ in the 2 heavier models. 

The abundances 
obtained near the age of the Hyades open cluster ranged from $-0.25$\,dex for the 
1.37\,\Msol{} model to $-1.25$\,dex for the 1.46\,\Msol{} model. Lithium 
is underabundant for
all models throughout evolution. 

By comparing these results with those shown in Fig.\,6 of \citet{richer93}, 
one first notices that the curves have very similar
behavior in time; the minima occur at nearly the same age and the curve shapes are
nearly identical. However, the underabundances obtained in 
our calculations are
systematically smaller. For
instance, their 1.43\,\Msol{} model is 250 times underabundant around 1.23\,Gyr,
while our 1.43\,\Msol{} model is only 14 times underabundant at the same age. A
careful analysis of the results determined that the difference was principally due
to the difference in the mass of the SCZ. If the convection zone
is homogeneous and emptied through its bottom, then:
\begin{equation}
\frac{d (X \Delta M_{\rm cz})}{d t}= Xw_i\rho
\end{equation}
where $X$ is the mass fraction in the SCZ (and on the surface), $\Delta M_{\rm cz}$
the mass in the SCZ, $w_i$ the diffusion velocity and $\rho$ the
density immediately below the SCZ. By integrating this equation and comparing
diffusion time
scales for the different values of $M_{\rm cz}$, 
the differences in surface lithium abundances for models of the same mass can be 
explained by the difference in $M_{\rm cz}$. This difference in CZ mass is 
mainly due to the inclusion of metal
diffusion in our models. In fact, the increased opacity in our models due to heavy metal
accumulation considerably increases opacity at the BSCZ, thereby extending it 
inwards.

\begin{figure}[!h]
\begin{center}
\includegraphics[scale=.92]{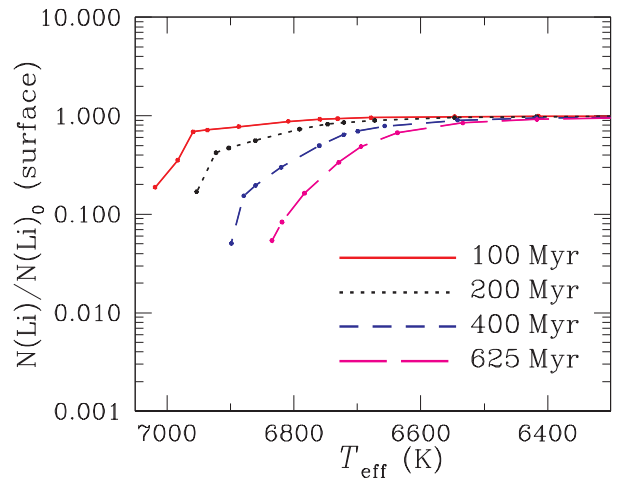}
\caption{The \teff-$X_{\rm Li}$ relation at different ages for the
models shown in Fig.\,\ref{fig:Lisurf}. Each point represents a calculated model. 
This plot can be compared directly to Fig.\,8
of \citet{richer93}.
}\label{fig:LiIso}
\end{center}
\end{figure}

In Fig.\,\ref{fig:LiIso},
the surface lithium abundances are shown for different isochrones of the models
shown in Fig.\,\ref{fig:Lisurf}. The gap position evolves toward 
cooler temperatures
with time. At 100\,Myr, only stars of 6900\,K have 
significant underabundances, while at 625\,Myr, stars as cool as 6700\,K have 
important
underabundances.

At 100, 200 and 400\,Myr, the gap obtained in the present calculations is 50-80\,K hotter than
the gap of the same metallicity shown in Fig.\,8 of \citet{richer93}. Our gap 
is also slightly deeper, as
the surface lithium on the hotter end of our isochrones keeps 
decreasing where surface lithium in the isochrones of \citet{richer93} start
increasing. This is a result of the smaller \gr(Li) obtained in our models (see 
Sect.\,\ref{sec:resultsLiBeB}, in particular
Fig.\,\ref{fig:gradLiBe}): the competition with Fe decreases \gr(Li) by a factor of
2, so that it doesn't reach $g$ in the hotter models of our isochrones. Therefore,
lithium is not supported in our diffusion only model. 

\subsubsection{Separated winds}
\label{sec:separation}
\begin{figure*}[!t]
\begin{center}                                                                                  
\includegraphics[scale=1.04]{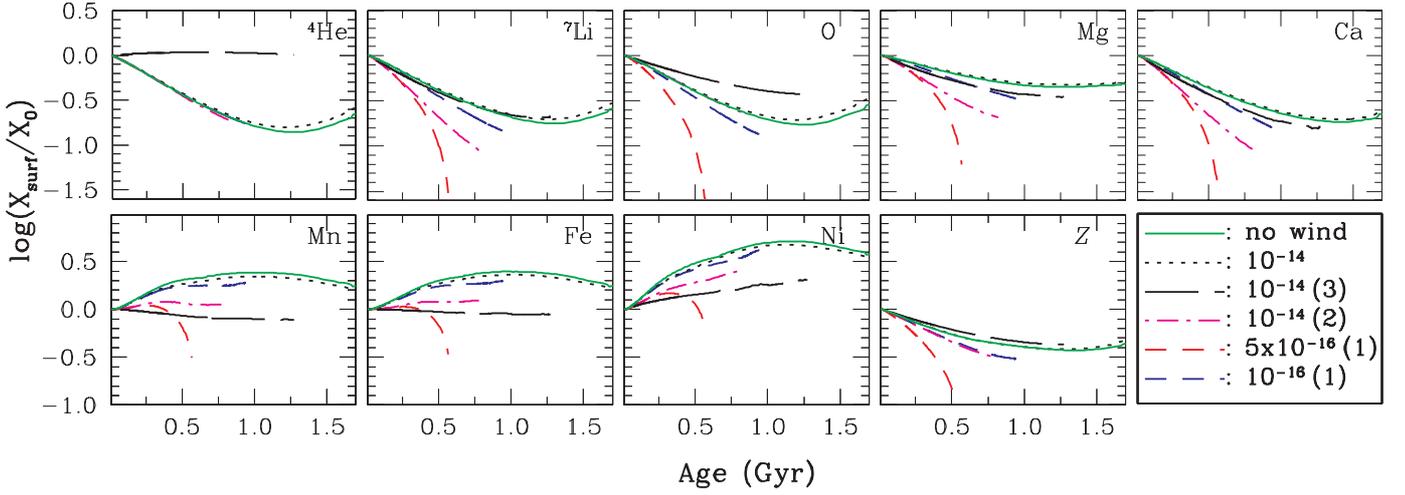}
\caption{Surface abundance evolution for selected elements of 1.40\,\Msol{} models with 
different types of wind solutions: one model with no mass loss (solid line), 
one model with a homogenous mass loss rate of 1.E-14\Mloss{}(dotted line), 
and 4 models with separated mass loss.  
For the separated winds, the configuration is indicated by 
a number (1, 2 or 3) which is explained
in Sect.\,\ref{sec:regimes} and in the text below.
}\label{fig:separation}
\end{center}
\end{figure*}
The effects of various separated wind
configurations on surface abundances of 1.4\,\Msol{} models are 
illustrated in Fig.\,\ref{fig:separation}. 
Along with the 1.40W1E-14 unseparated mass loss model and the pure diffusion model (no wind),  
three cases of separated winds were considered, as described in
Sect.\,\ref{sec:regimes}. 

In constrast to unseparated mass loss (dotted line) which, when comparing 
to the pure diffusion model
(solid line) in Fig.\,\ref{fig:separation}, 
{\it reduces surface anomalies for all elements}, 
separated mass loss can affect underabundant and overabundant 
elements differently. In case 1, 
for which all metals are ejected with the same relative concentration as in the atmosphere
whereas H and He remain bound, a 
mass loss rate of $1\times10^{-16}M_\odot/$yr\footnote{ 
This is smaller than the rate of metal depletion for the 
unseparated model 1.40W1E-14, since 
$Z\times {\dot M}= 1.999\times10^{-16}\Mloss$.} 
leads to $smaller$ overabundances and
$larger$ underabundances than unseparated mass loss. 
This is because in case 1, the internal wind term this mass loss leads to
is negligeable (see Eq. \ref{eqn:vwind}). Therefore, for 
underabundant elements, the surface 
convection zone is drained from its bottom (because the wind is too weak to 
support downward diffusing elements), 
as well as from its top (i.e. mass loss at the stellar surface). Overabundant 
elements, on the other hand, are depleted through the surface, while the 
bottom of the surface convection zone 
is replenished by atomic diffusion. When the 
mass loss rate is increased to $5\times10^{-16}\Mloss$ for case 1, the overabundances 
can rapidly evolve into underabundances since, without the wind 
providing sufficient 
replenishement from deeper inside the
star, the
elemental depletion at the surface quickly dominates  
the replenishement at the bottom of the convection zone. 
For the model with a mass loss rate of $5 \times 10^{-16}\Mloss{}$ 
underabundances of Li and Fe
reach $-1.5$\,dex and $-0.95$\,dex respectively at 570\,Myr, whereas 
for the $10^{-16}\Mloss{}$, the 
Li underabundance reaches $-0.55$\,dex around 625\,Myr, while 
Fe has an overabundane of 0.45\,dex. Generalized 
underabundances are attainable in the context of 
case 1.

In case 2, all low-FIP elements are depleted 4 times more 
rapidly than H and other high-FIP elements. 
Consequently, in comparison to the anomalies obtained for the 1.40W1E-14 model, the 
anomalies for the low-FIP elements (of which Li, Mg, Ca, Mn, Fe and Ni are shown) 
are greater for underabunt elements and smaller for overabundant elements, as in case 1. 
The Fe overabundance is reduced to 0.2\,dex around 625\,Myr, and the Li
underabundance reaches $-0.6$\,dex at the same age. However, for O, 
which is a high-FIP element, 
the curve is tucked in between the pure diffusion and unseparated mass 
loss curves (and is therefore impossible to see on the figure), which 
is what could be expected, since
the rate of depletion as seen by O, a high-FIP element, is smaller 
than $10^{-14}\Mloss{}$ because the lost mass 
has a higher concentration of low-FIP elements (see Sect.\,\ref{sec:regimes}).  

Finally, in case 3, for which H is included among the 
low-FIP elements, the results are quite
different: the depletion 
of H at the surface leads to an important
increase of He.  Since the He abundance at the surface remains 
slightly above its original value throughout the
simulation, the stellar structure is changed as the H-He SCZ is 
much deeper
on the \MS{} than for any of the other models. Around 625\,Myr, Fe is barely 
underabundant, while
Li is underabundant by $-0.5$\,dex\footnote{For a given element, [X/H] 
is about 0.1\,dex larger than $\log(X_{\rm surf}/ X_0)$ because of the 
surface underabundance of H.}.
At 625\,Myr, the model in case 3 is 100\,K hotter than 
the three other models at the same age.

\section{Comparison to turbulence models}
\label{sec:turb}
As previously mentioned, models with 
turbulence \citep{richer00,richard01,michaud05} 
have been quite successful at reproducing observed properties of AmFm stars 
both on the surface (i.e. abundance anomalies) and in the interior 
(pulsation properties of $\delta$ Scuti stars, see \citealt{turcotte00}).  
These models with turbulence put forth a scenario in which 
chemical separation takes 
place below 200\,000\,K, as opposed  to the scenario in which chemical 
separation immediately below the H convection zone is responsible 
for AmFm abundance anomalies (\citealt{watson71,alecian96}). 
The models calculated with mass loss offer an alternative scenario 
which is in fact a hybrid of 
both these scenarios. On the one hand, no external mixing is 
enforced outside of convection zones, which allows 
for chemical separation to occur throughout the stellar interior.
However, for most of the \MS{}
lifetime, the surface abundance is modulated by matter which is advected from deep inside
the star (while in turbulence models, it is mixed to that depth). 
In order to determine which of these scenarios must be favored, 
it is paramount to differentiate the models so as to enable observational tests. 

\begin{figure*}[!t]
\begin{center}
\includegraphics[scale=.42]{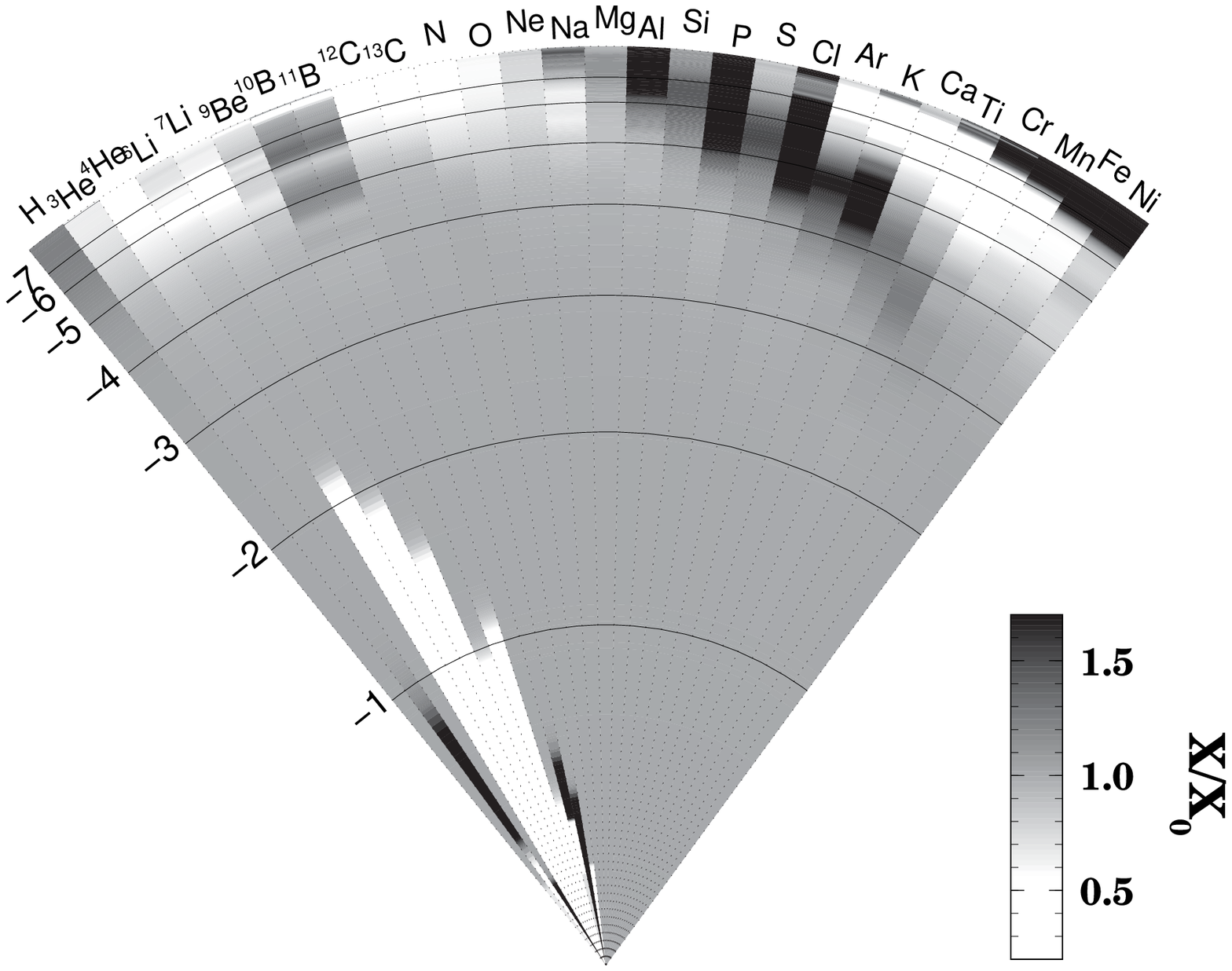}
\includegraphics[scale=.42]{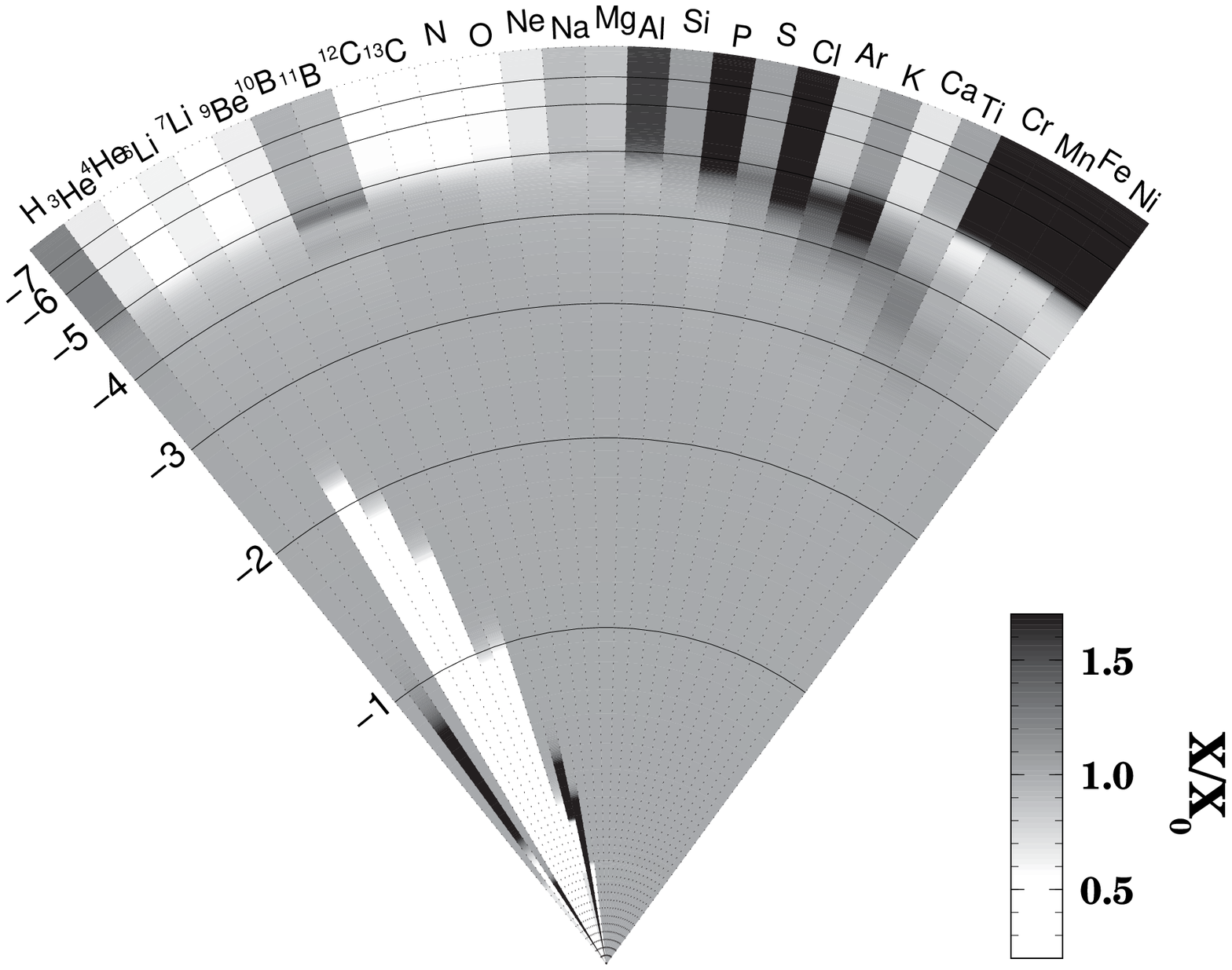}
\caption{Gray coded concentrations of two 2.50\,\Msol{} models at 500\,Myr. The scale 
of the radius is linear
although the value on the left hand side indicates $\DM$. The contrast is identified 
in the right insert and spans
the interval from 0.3 to 1.7 times the original elemental abundances. The anomalies 
affect the outer 25 \% of the 
radius. Nuclear effects appear for lighter elements near the stellar core. For the 
calculations shown in the left panel,
the competing process is mass loss while it is turbulence in the right panel. For
these two models, the
surface abundances are nearly the same (see Fig.\,\ref{fig:Sirius}).}\label{fig:eventails}
\end{center}
\end{figure*}

Even when surface abundances are quite similar in turbulent and mass loss models, 
the interior behaves differently. This can be seen by comparing the 
2.50W1E-13 and \hbox{2.50T5.2D1M-4} models in Fig.\,\ref{fig:eventails}
and Fig.\,\ref{fig:Sirius}. In Fig.\,\ref{fig:eventails}, which compares the internal 
concentrations of all
28 elements for a model with turbulent mixing 
and a model with mass loss, there is a 
stark contrast in the internal distribution of elements 
heavier than Ar, even though the 
surface abundances are quite similar for most atomic species 
(see Fig.\,\ref{fig:Sirius}). 
The  Fe abundance in the interval $\DM \simeq -6\,\,$to $-5$ is $4-5$ 
times larger in the turbulence model. The same can be
said for Ni which, between $\DM=-7\,$ and $-5$, differ by about a factor of 10.
This has an important effect on the local Rosseland opacity. 
These internal differences could therefore 
allow for some asteroseismological tests 
which could help
differentiate between the two models.

With respect to pulsations, one notes that around 250\,Myr, the 
age of Sirius (\citealt{liebert05}), 
the He\,II convection zone is still present in the 
turbulence model as well as in
the \hbox{2.50W1E-13} model, but not in the \hbox{2.50W5E-14} 
model (see Fig.\,\ref{fig:kabmu2.5}). As seen in the $\log \kappa_R$ panel,
opacities are quite similar in the He\,II convection zones 
for the \hbox{2.50W1E-13} and the turbulence
models, 
therefore, the mass loss model could be compatible with observed pulsation 
properties of $\delta$ Scuti stars (\citealt{turcotte00}). 
Further investigation 
is required to determine if the He\,II 
opacity bump in the 2.50W5E-14 model is sufficient 
to drive kappa mechanism oscillations. 

During evolution there are phases in which important surface 
abundance differences differentiate models with
mass loss from those with turbulence.  
For instance, Fig.\,\ref{fig:EffetAge} can be compared directly 
to Fig.\,15 of \citet{richer00}. 
The most glaring difference is
the surface behavior shortly after the onset of diffusion. In 
the turbulence model, 
abundance anomalies appear slowly
and gradually at the surface, whereas large anomalies appear 
at the surface as early as 5\,Myr in the
mass loss model. This is because all chemical separation that 
occurs near the surface, where timescales
are short, will rapidly be advected to the surface by 
the wind\footnote{
This will be discussed in further detail in a forthcoming 
paper which will look into the
effects of diffusion on the pre-\MS.} (see discussion in
Sect.\,\ref{sec:absurf2.5}). In the turbulence models, mixing is enforced
throughout the upper envelope, and effectively prevents any chemical 
separation near the surface.
Likewise, the abundances immediately following the turn-off are 
also different; while overabundant iron
peak elements such as Mn and Fe can become underabundant as the 
SCZ exposes regions in the envelope with
important gradients (as is the case at 524\,Myr), the internal 
variations in the 
turbulence context are much less important (as an
example, compare 
the Fe underabundance around $\DM\simeq-5$ in our Fig.\,\ref{fig:abint2.5} 
and Fig.\,4 of \citealt{richer00}), 
and so the variations at the surface will also be smaller. 
On the \MS{} however, the differences between the two models
are relatively quite small.

Similarly, for the \teff{} dependence, Fig.\,\ref{fig:EffetTeff} can be 
compared directly to Fig.\,16 of \citet{richer00}. 
The most glaring difference is the behaviour of iron peak elements. 
By comparing the models with masses between 1.7 and 2.3\,\Msol{} 
in both figures, one notes that 
in the turbulence regime, the iron peak surface 
distribution is the same for all \teff s at a given
age, while there is more significant variations for the same elements 
in the mass loss models within the same stellar mass interval. On the contrary, 
the lighter elements show less variations in the mass loss models in comparison
to the models with turbulence in the same mass interval. 



\section{Comparison to observations}
\label{sec:observations}
In order to constrain stellar models, and to determine whether 
turbulence or mass loss is the dominant macroscopic process 
reducing surface abundance anomalies, it is imperative to compare 
our results with observations. 
In order 
to carry out an accurate comparison 
one needs to constrain age, mass and 
initial composition, which as shown in the previous section, 
all affect surface abundances. To reduce the arbitrariness 
of the comparison, we chose three open cluster 
stars for which we have a good evaluation of the initial metal content 
as well as of the approximate age. We will also compare our results to the field star 
Sirius\,A and the binary system $o$\,Leonis. 

In the following sections, [N/H] has its usual meaning: 
\begin{equation}
\left[{\rm N/H}\right]=\log({\rm
N/H})_\star-\log({\rm
N/H})_\odot.
\end{equation}
As mentioned in Sect.\,\ref{sec:calcul}, this paper 
is part of a series of papers starting with \citet{turcotte98soleil}, 
where the 
mixing length used was calibrated 
using the Sun for given boundary conditions, as well as 
helium and metal content.  
For consistency, the same boundary condition, 
solar composition and mixing length are used for Pop I stars.  
Our models lead to abundance variations, or \textsl{anomalies}, 
relative to those original abundances.  A number of observers 
have similarly determined anomalies by differential methods 
with respect to solar abundances.  
Moreover, in the model atmospheres used for abundance 
determinations, most observers used the solar abundance mix either from 
\citet{anders89} or \citet{grevesse96}.
Because determinations 
were sometimes obtained with different solar 
photospheric abundances by different observers and 
sometimes with differential methods, 
their abundance determinations relative to 
the Sun (i.e. the anomalies) are used, when available, rather than absolute 
abundances. The uncertainty that inaccuracies in solar abundances 
lead to will be discussed in Sect.\,\ref{sec:conclusions}.

\subsection{Field Stars}
\begin{figure}[!t]
\begin{center}                                                
\includegraphics[scale=.53]{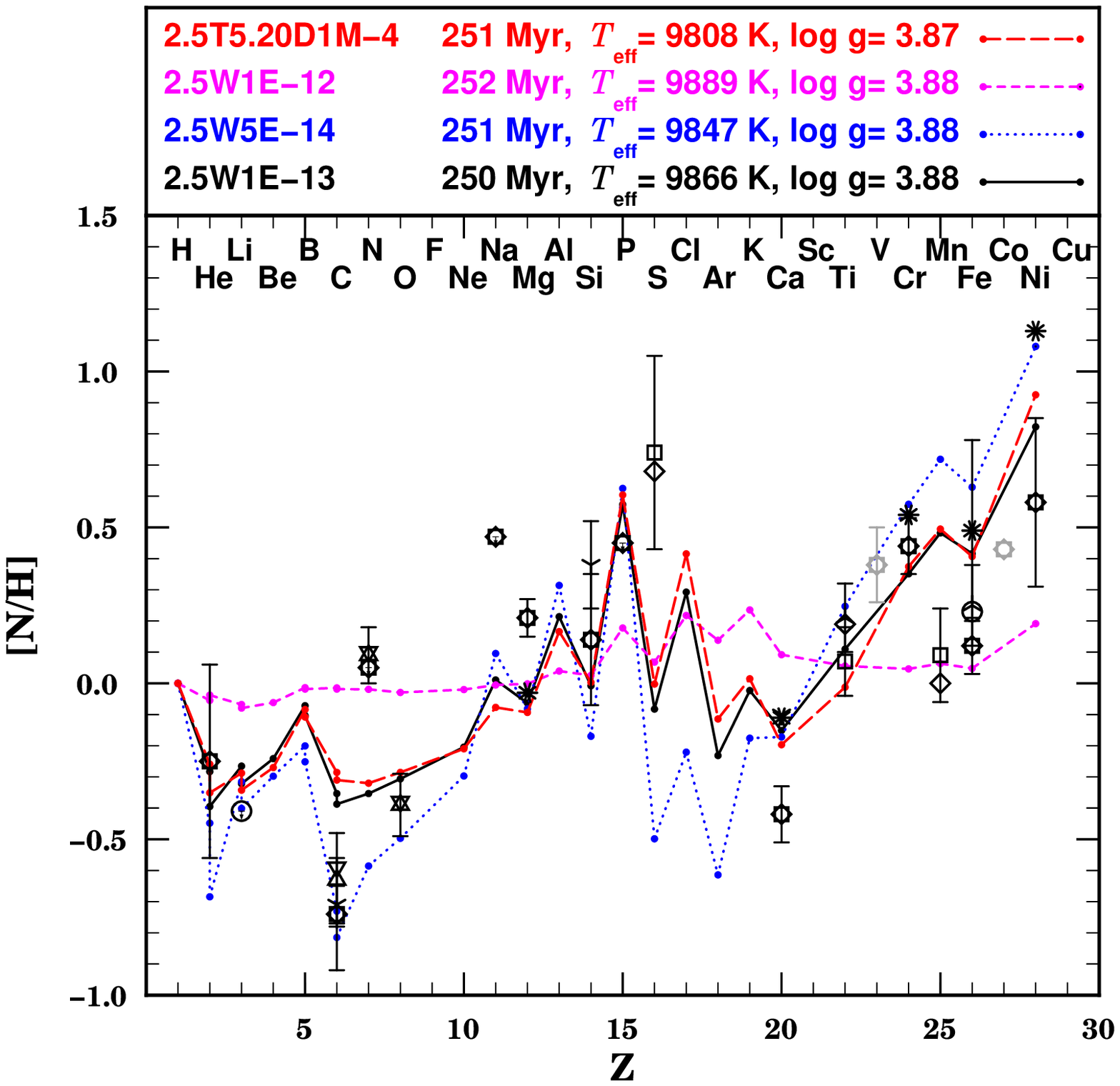}
\caption{Observations of the surface abundances of Sirius\,A 
(also known as $\alpha$ CMa, HR 2491 or HD 48915). 
{\it Circles}, \citet{burkhart91}; 
{\it Upright open triangles}, \citet{roby90}; 
{\it inverted open triangles}, \citet{lambert82}; 
{\it three-point stars}, \citet{lemke89};
{\it inverted three-point stars}, \citet{lemke90};
{\it squares}, \citet{hill95};
{\it diamonds}, \citet{hill93};
{\it asterisks}, \citet{hui-bon-hoa97}.
Calculated values are shown for 3 models with 
mass loss (2.50W5E-14, 2.50W1E-13 an 2.501E-12), 
as well as the model with turbulence (2.50T5.2D1M-4) which represents 
the best fit from Fig.\,18 of \citet{richer00}. 
All models were calculated with a solar ($Z=0.02$) 
initial metallicity and are shown at an age of 250\,Myr. 
Vanadium, scandium and cobalt are grayed out since they 
are not included in our calculations. The internal abundance distributions 
of the 2.50W1E-13 and 2.50T5.2D1M-4 models 
are shown in Fig.\,\ref{fig:eventails}.}\label{fig:Sirius}
\end{center}
\end{figure}

Sirius\,A is the most studied hot 
Am star ($\teff \simeq 9800\,$K), with a mass of about 2.14\,\Msol{} 
and an age of approximately 250\,Myr (\citealt{gatewood78}; 
see also Sect.\,4.1 of \citealt{richer00}).
Figure\,\ref{fig:Sirius} shows surface abundance determinations  
for 19 chemical species (16 of which are included in our calculations) 
from 8 different 
papers.  

First, for most elements, there is considerable 
scatter among observers. For instance, there is   
a 
0.3\,dex difference in Si abundance, as well as a 
0.4\,dex difference in Fe abundance,
which is the most carefully determined element.
We compared this data to 3 models with mass loss 
(2.50W1E-12, 2.50W1E-13 and 2.50W5E-14) as well as to
the model with turbulence from \citet{richer00} which best reproduced the 
data (2.50T5.2D1M-4). Of the 16
observed elements which are included in our calculations, 
12 (He, Li, O, Na, Mg, Si, P, Ca, Ti, Cr, Fe and Ni) are 
well reproduced by both the turbulence
model and the 2.50W1E-13 model. 
It is interesting to note that  
\citet{bertin95} determined, from Mg\,II lines, that the observed 
mass loss rate of Sirius\,A is between $1.5 \times 10^{-12}$
and $2 \times 10^{-13}$. 
The overall fit is also just as good for the 2.50W5E-14 
model since it is also able to 
reproduce the carbon abundance, though it perhaps overevaluates the iron peak
abundances.   
The S abundance
is not at all reproduced by our calculations; however, the 
observer gives little credibility to its value
\citep{hill95}. It is clear that the model with a mass 
loss rate of $10^{-12}$ \Mloss
does not lead to the observed surface abundance
pattern. Finally, surface abundance observations are not sufficiently accurate 
to enable a differentiation between
turbulence and mass loss models.

In \citet{michaud05}, the binary system $o$\,Leonis (HD 83808/83809) has been 
interpreted as consisting 
of two AmFm stars with masses of 2.12 and 1.87\,\Msol\,\citep{griffin02}. The authors 
show that two models with turbulent mixing, of 2.24 and 1.97\,\Msol{} respectively, are
able to reproduce the observed features (namely both positions in the H-R 
diagram as well as surface abundances)
of the A and B components of the binary system within the observational error bars.
\begin{figure}[!t]
\begin{center}                                                                                               
\includegraphics[scale=.89]{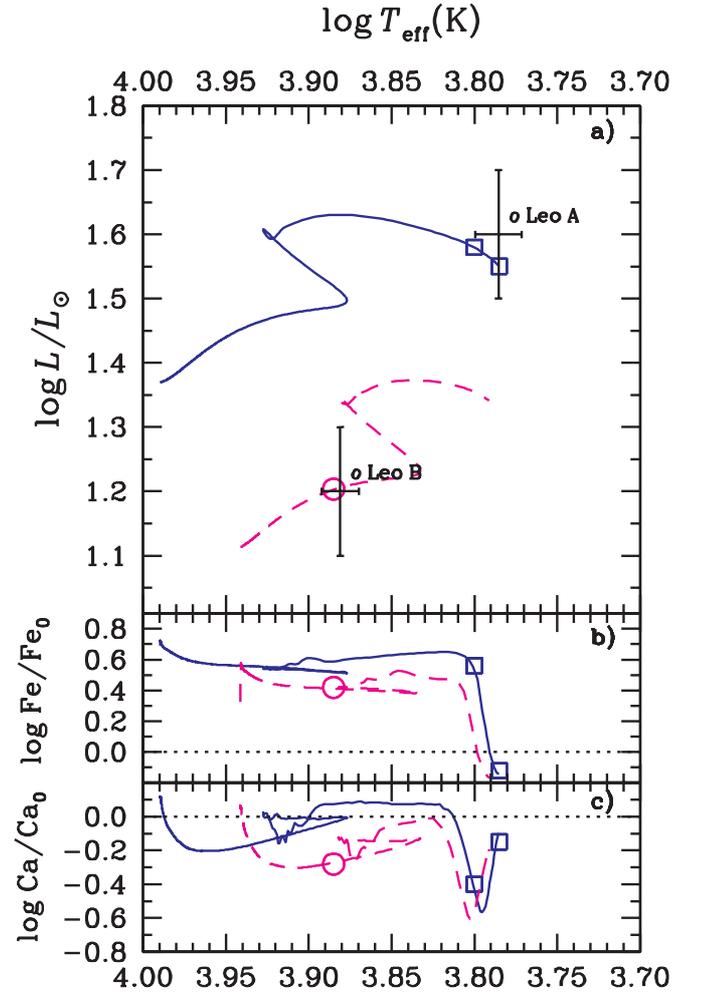}
\caption{A model of 2.2\,\Msol{} with a mass loss rate of $5 \times 10^{-14} \Mloss$ 
(solid line) and a model of 1.9\,\Msol{} with the same mass loss rate (dashed line) are plotted 
in an H-R diagram (a). The
observed position in the H-R diagram of the primary and secondary components 
of $o$\,Leonis (with error bars) are shown with crosses. 
The squares indicate two possible positions on the primary's evolution path which 
fall within observational error bars. The circle
indicates the position of the secondary at the age of the primary indicated 
by the square. 
The surface values of Fe and Ca are shown relative to their original values in (b) and (c).
}\label{fig:oLeo}
\end{center}
\end{figure}
Similarly, models of 2.20 and 1.90\,\Msol{} with a mass loss 
rate of $5 \times 10^{-14}$\Mloss\, are 
able to reproduce 
the H-R position of both components. At the age indicated
by the squares (750\,Myr and 750.5\,Myr), the A component is in the rapid evolution 
stage that follows the depletion of hydrogen in the core (the luminosity change 
is related to the star's adjustement to H-shell burning). At this age,
the B component is still in the slowly evolving MS stage,
which explains why both circles overlap on the graph.  

The two squares in Fig.\,\ref{fig:oLeo} show that the Fe abundance 
can vary between $-0.2$\,dex and 0.6\,dex within the \teff{} error bar for the 
A component, while the 
corresponding Ca surface abundance varies from $-0.18$ to $-0.4$. Throughout
most of this interval, the 2.20\,\Msol{} model has anomalies which are
typical of Am stars (Fe overabundance coupled with a Ca underabundance). 
For either of the values for the A component, 
the 1.90\,\Msol{} model has
an overabundance of Fe of 0.4\,dex coupled with 
an underabundance of Ca which attains $-0.3$\,dex, 
both typical Am star anomalies as well.

Both the turbulent model of \citet{michaud05} and the mass loss model can reproduce 
the AmFm character of components A and B. While the AmFm character of component A can be 
fitted by the turbulent model for its exact observed \teff{}, the mass 
loss model only generates 
typical AmFm iron overabundances for a part of the error bar on 
the hot side of the observed \teff.


\subsection{Open Cluster Stars}
\label{sec:opencluster}
\begin{figure}[!t]
\begin{center}                                                                        
\includegraphics[scale=.53]{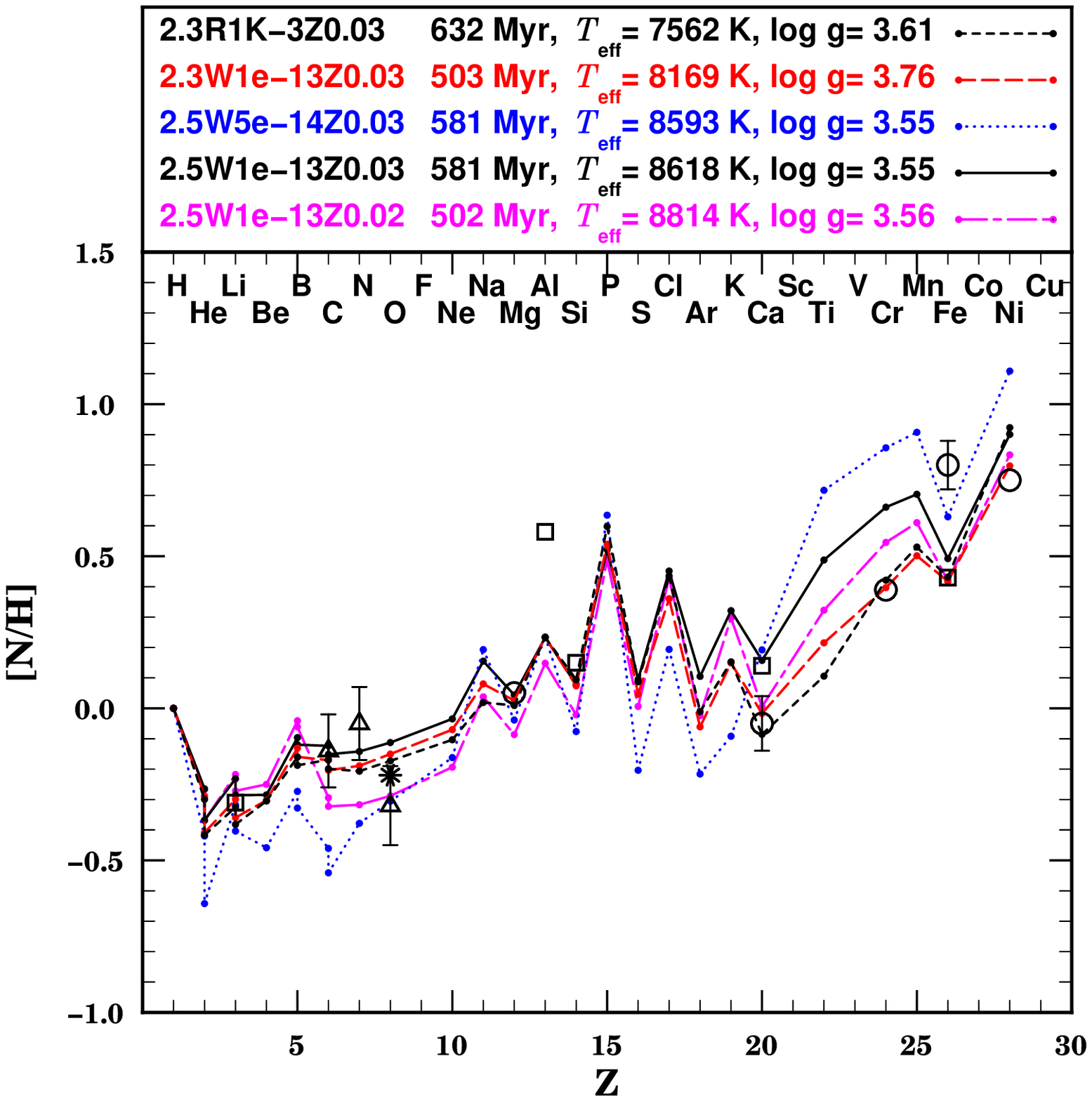}
\caption{Observed surface abundances of 68 Tau 
(also known as vB 56, HR 1389 or HD 27962), the hottest 
star (blue straggler) from the Hyades open cluster. 
{\it Circles}, \citet{hui-bon-hoa98};
{\it triangles}, \citet{roby90};
{\it squares}, \citet{burkhart89};
{\it asterisks}, \citet{takeda97}.
communication. 
Calculated values are shown for 4 models with varying 
mass loss rates as well as
the model with turbulence which best reproduced 
the data (2.30R1K-3Z0.03, see \citealt{richer00}). 
One model was calculated with $Z_0=0.02$, while all other models 
were calculated with an initial metallicity of $Z_0=0.03$. Metallicity 
is indicated in the model name.
}\label{fig:68tau}
\end{center}
\end{figure}

In Fig.\,\ref{fig:68tau}, we compare our results
to observed abundance determinations for the hot Am star 68\,Tau 
($\teff\simeq9050$\,K, \citealt{netopil08})
from the Hyades open cluster. The cluster age has been quoted between 
625\,Myr (\citealt{perryman98}) 
and 783\,Myr by
\citet{varenne99}. Its quoted metallicity has also ranged between 
$Z_0 \simeq 0.024$ (\citealt{perryman98,gratton00})
\footnote{Using the Hyades iron enrichment factor of \citet{gratton00} 
to multiply the metallicity determined by
\citet{asplund09} for the solar mixture, 
the Hyades metallicity becomes
$Z_0 \simeq 0.019$. The actual value would
likely lie somewhere between 0.019 and 0.03. The curve with $Z_0=0.02$ in Fig.\,\ref{fig:68tau} illustrates 
the impact of uncertainties.} 
and $Z_0=0.03$
(\citealt{cayrel85}) using F and G star iron abundances as indicators. 
In order to 
reflect this
metallicity, the selected models have been calculated 
with an initial  metallicity of $Z_0=0.03$, which was also used in models
from \citet{richer00}\footnote{Of course, 
[N/H] at $t=0$ for models with $Z_0=0.03$ is above solar for all metals.}. 
We 
have attempted to make a
compromise between fitting age and 
\teff: three models have a mass of 2.50\,\Msol{} and
one has a mass of
2.30\,\Msol{}. The 2.50W1E-13Z0.03 and 2.50W5E-14Z0.03 
models are 
on a short \teff{} upswing which arises as hydrogen nears 
depletion in the core (in Fig.\,\ref{fig:oLeo} for instance, 
it is the segment which immediately follows the \MS{}, 
spanning from $\log \teff \simeq 3.87$ 
at its bottom to $\log \teff \simeq 3.93$ at its top). 
In terms of stellar age, this upswing only lasts 3\,Myr before the star 
starts its descent onto the red giant
branch. While it has very little effect on surface abundances, models were 
chosen at this age in order 
to be closer to the star's 
surface temperature. Given the large spread in abundances between observers,
the fit is almost perfect with the 
2.30W1E-13Z0.03 model, which is slightly cooler and younger, 
yet is still on the \MS{}. 
Of 15 observed elements, only 
Na, Al and Mn (arguably just Al) are not reproduced. 
The fit is as good if not better (because of Ni) than the fit 
obtained with the model with turbulence. 
The 2.50W1e-13Z0.02 model was  added in order to 
illustrate the effect of reducing initial metallicity on absolute abundances 
(see also Fig.\,\ref{fig:EffetZ}). The fit with observations is 
better than for the 2.50W1-13Z0.03 model, since 
the iron peak abundances with respect to solar abundances are smaller 
in the lower metallicity model. By comparing these two
curves one can conclude that, for a given mass loss rate, 
a 0.18\,dex reduction of initial metallicity can, at most, lead to a 
0.18\,dex reduction for elements which are not supported, 
such as  C, N and O, and a 0.09\,dex reduction for elements which are supported such as Fe. 
See also Sect.\,\ref{sec:Effets}.  

\begin{figure}[!t]
\begin{center}                                                                                               
\includegraphics[scale=.53]{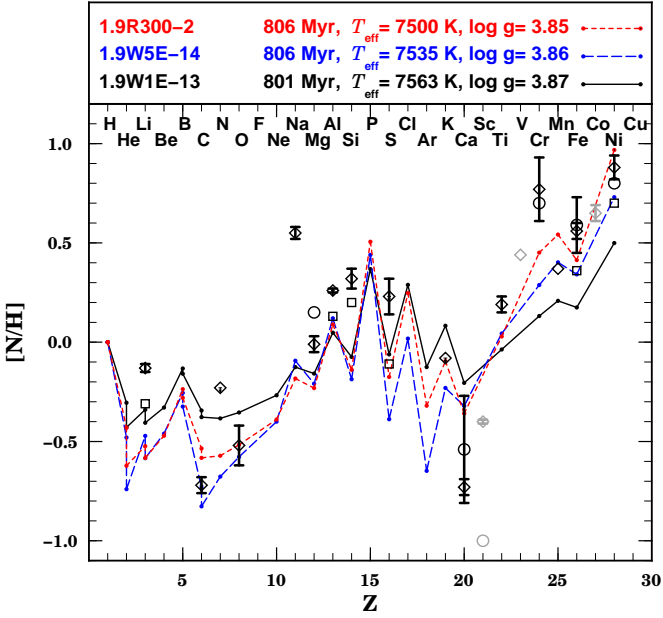}
\caption{Observed surface abundances of HD 73045 ($\teff\simeq$ 7500\,K) of the 
Praesepe open cluster which has an approximate age of 800\,Myr.
{\it Circles}, \citet{hui-bon-hoa98};
{\it squares}, \citet{burkhart00};
{\it diamonds}, \citet{fossati07}. Curves correspond to models listed 
at the top of the figure. All models were computed with solar initial abundances ($Z_0=0.02$).
}\label{fig:praesepe}
\end{center}
\end{figure}
In Fig.\,\ref{fig:praesepe}, we compare 2 models of 1.9\Msol{} with mass loss 
as well as a model with turbulence to the
observations of the star HD 73045 ($\teff\simeq7500\,$K) from the 
Praesepe open cluster which 
has an approximate age of 800\,Myr and a solar metallicity.
There are 15 observed elements which can be compared to our simulations, 
although 3 determinations (N, K and Mn), result from a single line and 
therefore could be inacurrate. 
Again, note the large discrepancies between observers. Only the 1.90W5E-14 and 
the turbulence model
can reproduce either the overabundant iron peak elements 
or the underabundances of C and O.
The abundances of Na and Si are not reproduced by either of the models.

\begin{figure}[!t]
\begin{center}                                                                                               
\includegraphics[scale=.53]{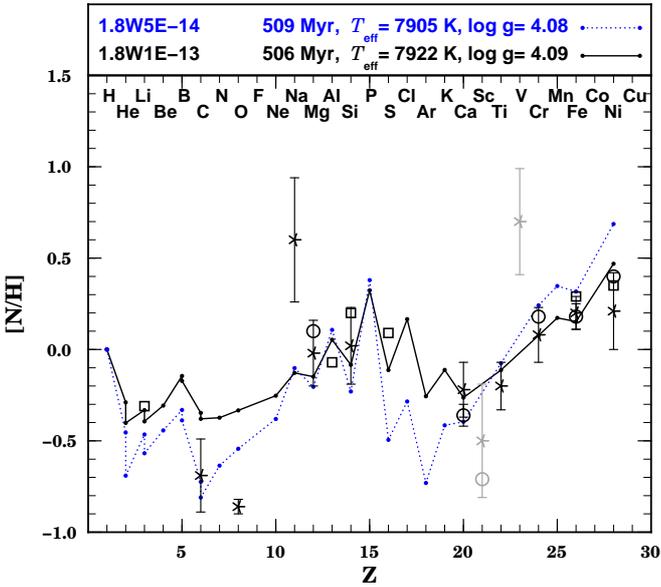}
\caption{Observed surface abundances of HD 108486 ($\teff\simeq$ 8180\,K) of the 
Coma open cluster which has an approximate age of 500\,Myr.
{\it Circles}, \citet{hui-bon-hoa98};
{\it squares}, \citet{burkhart00};
{\it five-point stars}, \citet{gebran08coma}. Curves correspond to models listed 
at the top of the figure. All models were computed with solar initial abundances ($Z_0=0.02$).
}\label{fig:coma}
\end{center}
\end{figure}

Finally, we have compared our models to observations of the 
Coma Berenices star HD 108486 (Fig.\,\ref{fig:coma}). Coma Berenices is
an open cluster with an age of about 500\,Myr and with a metallicity which is 
about solar. 
We have matched the star's \teff{} and age 
quite well with two 1.8\,\Msol{} models with mass loss. 
Except for O and Na, which are not reproduced by any of the two 
models, most elements are fitted by both models. 
Assuming error bars for 
S and
Al which are similar to those for other elements, we can state that 10 of the 13 abundances 
can be reproduced by the 1.80W1E-13 model, and 9 by the 1.80W5E-14 model.

Although we have opted not to add any extra figures, our results are also compatible 
with observations of Ca overabundances (see Fig.\,\ref{fig:alecian})
in very 
young open clusters such as the Pleiades 
\citep{gebran08pleiades,hui-bon-hoa98} and $\alpha\,$Persei \citep{hui-bon-hoa99}. This 
is noteworthy
since models with turbulence do not predict such an overabundance 
(see Figs.\,10 of \citealt{richer00}).

\subsubsection{Lithium gap}
\label{sec:LiGap}
\begin{figure}[!t]
\begin{center}                                                                                               
\includegraphics[scale=.98]{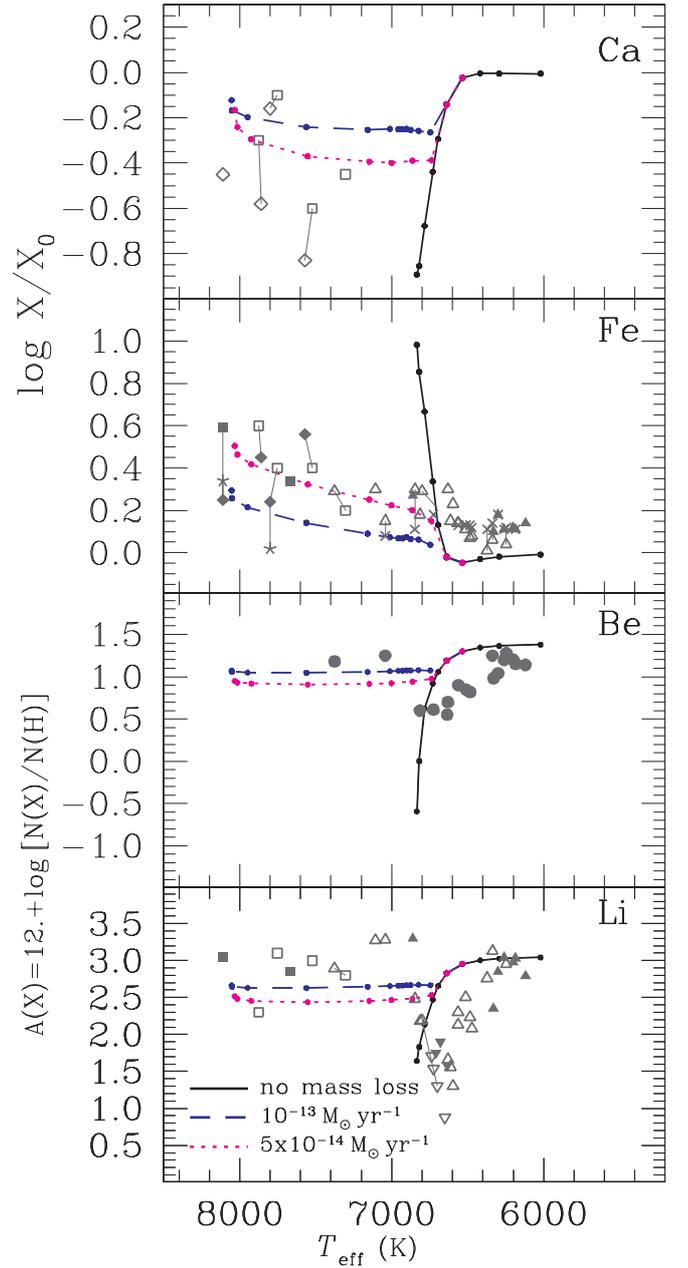}
\caption{Lithium, beryllium, iron and calcium abundances for models with and without mass loss
at 625\,Myr, the approximate age of
the Hyades open cluster. All models were calculated with 
an initial metallicity of $Z_0=0.02$ and the original Li abundance was set at N(Li)=3.05. The Li observations
are from ($\triangle$, and $\triangledown\,$ for upper-limits) \citet{boesgaard86}, 
($\blacktriangle$ and $\blacktriangledown$) \citet{boesgaard88},
($\square$) \citet{burkhart89}  and
($\blacksquare$) \citet{burkhart00}. Be abundances are taken from ($\bullet$) \citet{boesgaard02}. 
Additional calcium and iron abundances 
are also shown for all stars with lithium determinations: ($\times$) \citealt{boesgaard90};
($\star$) \citealt{takeda97}; ($\diamond$) \citealt{hui-bon-hoa98}
. All
stars with multiple determinations are connected by a line segments. Calculated models are
indicated by dots along the curves.}
\label{fig:LiGap}
\end{center}
\end{figure}
In Fig.\,\ref{fig:LiGap}, models with and without mass loss are compared to 
lithium, beryllium, calcium and iron observations in and around
the Hyades lithium gap. Lithium determinations are shown for  
F stars and AmFm stars (normal A stars and other peculiar A stars are 
omitted). Iron and calcium abundances are shown only for
stars which had a lithium determination. All beryllium abundances for F stars in
\citet{boesgaard02} are shown\footnote{Lithium abundances
from \citet{boesgaard86} and \citet{boesgaard88} are prefered over the revised
values from \citet{boesgaard02} because there is a greater 
number of stars. Nonetheless, in the
more recent paper, $all$ previous determinations are 
revised upwards by 0.09 to 0.4\,dex.}. 
When multiple 
observations for the same star were
available, the different determinations are connected by a line segment. 
This 
gives an evaluation 
of the uncertainty. 
All models calculated with 
mass loss rates of $5\times 10^{-14}\Mloss$ and 
$1\times 10^{-13}\Mloss$ which were still on the \MS{} at 625\,Myr are 
shown. Models with a mass loss rate of 
$1\times 10^{-14}\Mloss{}$ are 
omitted since they result in surface abundances which are very 
similar to diffusion only models 
(compare the diffusion only and 1.40W1E-14 models in 
Fig.\,\ref{fig:separation}). Therefore in the following
discussion, results from diffusion only models can be assimilated 
to models with unseparated mass 
loss $\leq 1\times 10^{-14}\Mloss{}$.
All models were 
calculated with an initial metallicity of $Z_0=0.02$, 
although the metallicity of the Hyades is above 
solar ($Z=0.024$, see Sect.\,\ref{sec:opencluster}).  
The original 
value of lithium was set to A(Li)=3.05. This value fits lithium 
determinations for the stars at the top of 
the cold side of the gap, which, since diffusion plays only a 
small role for these stars, 
probably reflect
the cluster's original Li content (unless there 
is significant pre-\MS{} burning). 
Following the
same logic, 
A(Be) was set to 1.40. 
  
According to our calculations, atomic diffusion in the absence 
of competing processes 
leads to an important 
reduction of surface lithium abundance. The smallest 
lithium anomaly 
was obtained for the 1.10\,\Msol{}
model, for which surface lithium was reduced by 0.015\,dex at 625\,Myr. The 
largest lithium 
reduction at 625\,Myr is by about $-1.4$\,dex (or a factor of 25), for the
1.46\,\Msol{} model without mass loss. This does not quite
reach the bottom of the 
gap of \citet{boesgaard86}\footnote{It has been suggested to revise these observations 
upwards by up to 0.4 dex (see the discussion in Sect.\,2.1.2 
of \citealt{michaud91} and \citealt{boesgaard02}). The maximum depletion encountered 
in the deepest part of the gap might so be closer to a factor of 50.}.
As seen in Fig.\,\ref{fig:Lisurf}, the 
1.47\,\Msol{} model without 
mass loss would likely have reached a lower lithium 
abundance had it been able to converge up to the age of the Hyades, thus 
reconciling some of the difference. In fact,
the mass of the model which would have attained the 
lithium gap minimum can be 
deduced from Fig.\,\ref{fig:gradLiBe}.
Because Li is not supported until $\gr \simeq g$ just below the surface 
convection zone at 625\,Myr, the heaviest model for which 
the BSCZ is located where 
$\log T\gtrsim 5.4$ throughout its evolution will
represent the gap minimum, since it is for this model that Li 
is sinking fastest.
From Fig.\,\ref{fig:gradLiBe}, while the 1.55\,\Msol{} model is clearly on the hot side and the 1.43\,\Msol{}
model on the cold side, it is the 1.46\,\Msol{} model that should be closest to the bottom of the gap. 
Furthermore, if the competition 
with Fe 
that is prescribed by the OPAL opacities and calculated in these 
models without mass loss
is correct, then diffusion alone cannot explain the increase of 
Li on the hot side of the gap.
The \gr(Li) in the atomic diffusion 
only models will remain 
smaller than $g$ by a factor of at least 2. However, given the uncertainties 
discussed in Sect.\,\ref{sec:methods} on the location of Fe lines, \gr(Li) 
could very well attain $g$ near $\log T \simeq 5.3$, in which case lithium 
would be supported, and would consequently exhibit different surface behavior.
    
For the models without mass loss on the cold side of the gap, 
the Fe abundances are in agreement with 
observations up to about 6800\,K, after which the 
calculated overabundances become too large. 
The discrepancy between the Fe curve and the observations for 
$\teff<6700\,$K is related to our models having a solar initial metallicity,
whereas the Hyades stars were formed in a metal 
rich environment.

The models with mass loss rates of $5 \times 10^{-14}\Mloss{}$ and $10^{-13}\Mloss{}$ 
cannot explain the depletion encountered within 
the gap. However, given reasonable error bars, they
are consistent with the almost constant lithium and beryllium abundances observed 
for $\teff>7200\,$K. Both mass loss rates 
lead to models which reproduce the observed Fe abundances 
between $ 6000 \leq \teff \leq 8200\,$K and, in particular,  
the increase in Fe abundance for $\teff>7200\,$K, 
which is compatible with the 
AmFm character of these stars.
Given the large discrepencies in determinations, most Ca 
abundances are also compatible with the models with mass loss.  

Neither the calculated gap minimum nor the shoulder on the cold 
side of the gap match the observed 
position in \teff. 
By comparison to Fig.\,9 of \citet{richer93}, their calculated 
depth for the gap ($-1.6$\,dex) resembles
the depth obtained in our calculations ($-1.4$\,dex, see 
discussion in Sect.\,\ref{sec:lithium}). 
The 
shoulder on the cold side of the gap obtained in the present calculations matches 
the curve they obtained
with $Z_0=0.02$ within $\pm 50\,$K. Accordingly, as also seen in this same Fig.\,9, if we had 
chosen $Z_0=0.03$ ([Fe/H]=+0.18), 
some of the 150\,K difference 
would have been recuperated as the gap minimum would have been shifted 
toward cooler
temperatures by 50-80\,K. There is also a $\pm 50-100\,$K 
uncertainty on the observed potition of the gap (see discussion 
in Sect.\,2.1.2 of \citealt{michaud91}). 
The uncertainty on the age of the 
Hyades (from 625 to 783\,Myr, see Sect.\,\ref{sec:opencluster}) could 
also 
account for some of the difference as 
illustrated in Fig.\,\ref{fig:LiIso}. 
As the age of the isochrones increases, the \teff{} at which 
lithium abundances fall off also decreases. 
The real problem in explaining the Li gap with atomic diffusion is not with the exact \teff{}
of the gap nor its depth, but rather with the calculated Fe overabundances which are
not observed, and the related difficulty in calculating \gr(Li) on the hot side of the 
gap due to Fe lines.

\section{General discussion and conclusion}
\label{sec:conclusions}
\subsection{Summary of results}

Evolutionary models including both atomic diffusion and unseparated mass loss  
explain the main abundance anomalies of AmFm stars (Sect.\,\ref{sec:opencluster}). When
mass loss is assumed to be the only macroscopic process
competing with atomic diffusion, observed abundance 
anomalies from open cluster stars as well as Sirius\,A and $o$\,Leonis 
constrain mass loss rates 
to $2-5$ times the solar mass loss rate. As shown in Sects.\,7.1 and 7.2, 
models involving
mass loss are as capable as models involving turbulence in 
explaining observations of AmFm stars. 
This is because in both instances, the important separation occurs at the same depth 
($\Dm\simeq 10^{-6}-10^{-5}$) for most of the \MS{} life.
Whether the mass loss model is to be preferred over the turbulence model
is difficult to assess given the large observational uncertainties. 
However, as shown in
Fig.\,\ref{fig:eventails}, the internal distribution of 
elements is different between the two cases for most
elements. With differences reaching a factor of $4-5$ for abundant 
elements such as Fe, there should 
be effects on local opacities and thus on pulsations. 
Asteroseismic tests could perhaps distinguish 
between the two\footnote{\citet{carrier07} did not detect 
pulsations which could have been a signature of iron accumulation in the Am star 
HD209625.}.

In the mass loss regime, chemical separation affects up to $10^{-5}$\Msol{}
of a star's mass or, equivalently, 20 to 25\% of the stellar radius (see 
Fig.\,\ref{fig:abint2.5} and Fig.\,\ref{fig:eventails}). 
For any given element, 
as long as the wind velocity is greater in amplitude than the downward settling velocity, 
the local abundance solution is determined by 
flux conservation; local abundances 
adjust as the flux quickly becomes constant throughout the outer envelope (see Fig.\,\ref{fig:flux}). 
As a result, 
the surface abundances depend on matter which is advected from deep 
inside the star (see discussion in Sect.\,\ref{sec:absurf2.5}). This differs from the 
models of \citet{watson71} and \citet{alecian96} in which surface abundances 
depend on the outer $10^{-10}\Msol{}$. This can also be contrasted to the solution obtained
in the models with weak or $fully$ $separated$ mass loss 
presented in Sects.\,\ref{sec:abint1.5} and \,\ref{sec:separation}. 

When flux is conserved throughout the envelope, abundance gradients 
which form near the surface, between 
surface convections zones for example, 
have no
effect on the surface solution once the star has
arrived on the \MS{} (see discussion in Sects.\,\ref{sec:absurf2.5} 
and \ref{sec:Effets}). In this instance, if the age of interest is 
greater than $M_{\rm BSCZ}/{\dot M}$, where $M_{{\rm BSCZ}}$ is 
the mass above the bottom of the deepest surface
convection zone, one can obtain a nearly similar surface 
solution by approximating that
abundances are homogeneous from the surface
to the bottom of the deepest SCZ. 
However, early in the evolution, only matter from superficial 
layers has had time to be advected,
and thus surface abundances obtained here depend on separation that occured 
close to the surface as first studied for
Ca by \citet[see also Sect.\,\ref{sec:surfab1.5}]{alecian96} and confirmed 
observationally (see end of Sect.\,\ref{sec:opencluster}).
This favors models involving mass loss rather than turbulence. 
Likewise, variations 
obtained near the surface, which do not appear in models with turbulence, 
have an effect on the PMS (Fig.\,\ref{fig:EffetAge}) and  
will be discussed in
a forthcoming paper.  

In all models heavier than 1.47\,\Msol{} without mass loss or with an unseparated 
mass loss rate $\leq 1\times10^{-14}\Mloss{}$, the accumulation of Fe and Ni 
around $T=200\,000\,$K leads to the appearance of a thin radiative layer 
which separates the
iron peak convection zone from the surface H-He convection zone. This accumulation forms
before the appearance of a small inversion of the local molecular weight gradient
inversion
(see Sect.\,\ref{sec:abint1.5}). The inclusion of thermohaline convection
as suggested by \citet{theado09} could have an effect 
on abundances in the region, though convection occurs 
even when there is no molecular weight gradient inversion. 
This would require further investigation. Nonetheless,
the appearance (or not) of the iron peak convection zone does 
not have a significant effect on the
surface solution, nor does it appear in models with mass loss which 
adequately reproduce observed
abundance anomalies of AmFm stars (Sect.\,\ref{sec:observations}). 

Since this paper is
a part of a series which explores the various macroscopic processes which compete
with atomic diffusion in AmFm stars, it is important that the models 
be as similar as possible to those used in previous calculations (e.g. those with turbulence)
in order
to isolate the effects due specifically to mass loss.   This is one of the
primary motivations for using the same initial solar abundances as in 
previous calculations,
rather than the revised \citet{asplund05,asplund09} abundances 
(see also the discussion
in Sect.\ref{sec:calcul}). Varying too many
things at once could obscure results and introduce further uncertainty. 
Furthermore, there is a controversy on solar abundances, 
since heliosismology strongly favors the older \citep{grevesse96} over the newer 
\citep{asplund05,asplund09}
composition. One may then view the abundance 
differences between the two sets as an evaluation of uncertainty. 
Since solar abundances are used throughout this paper, the uncertainty 
on solar abundances leads to uncertainties on the absolute values of 
all abundances.  
As shown in Fig.\,\ref{fig:68tau}, a factor of 1.5 reduction (or 0.18\,dex) 
of the
original $Z$ leads to 
a similar reduction (0.18\,dex) of the expected abundances 
of atomic species that are not supported, as well as a smaller reduction of 
0.09\,dex for species such as \Fe{} which are supported by \gr{}. The 
fit for the 
abundances of 68\,Tau is about the same for both values of $Z$ as seen 
in Fig.\,\ref{fig:68tau}.  
Equivalently, compensating the change of \Fe{} abundance would require 
reducing the mass loss rate from 
 $10^{-13} \Mloss$ to  $7 \times 10^{-14} \Mloss$ according to 
 the results shown on Fig.\,\ref{fig:absurf}.  
This may be viewed as the uncertainty on the mass loss rate resulting from the uncertainty of 
solar abundances\footnote{To significantly 
improve the evaluation of the effect of changing  to another set of solar 
abundances would require first recalibrating
the mixing lenght using a solar model,
then carrying out calculations for AmFm stars 
for both turbulence and mass loss, as well as 
reanalyzing observations of AmFm star abundances using the new solar abundances. 
This is  outside the scope of the present
paper.}.

\subsection{Further implications}

Atomic diffusion alone $cannot$ explain all 
characteristics of the Hyades 
lithium gap, nor can unseparated mass
loss. 
The cold 
side of the gap can only be
reproduced by diffusion only models or 
models with ${\dot M}\leq 1\times 10^{-14}\Mloss{}$, whereas the hot
side of the gap and the AmFm character of stars 
for $\teff \geq 7200\,$K $require$ 
a stronger mass loss rate. Moreover, separated mass loss (see 
Sects.\,\ref{sec:separation} and \,\ref{sec:LiGap}) 
seems required to explain observed Li underabundances 
near the bottom of
the gap as well as reduce the calculated Fe overabundances. In 
Fig.\,\ref{fig:separation}, in comparison to the \emph{diffusion only} model, 
the curve for case 2 shows both larger underabundances 
of Li as well as smaller overabundances
of Fe. Well tuned \emph{fully separated mass loss} (case 1) could do the same. 
Similarly, the model for case 3 has a nearly flat 
Fe surface abundance coupled with 
similar Li underabundances to the \emph{diffusion only} model. 
It does not seem justified to 
further speculate on the role of separated winds in Li gap stars
until we have a better understanding of separation mecanisms within stellar winds.
Since radiative forces
generally increase with \teff, the above mentioned 
increase in mass loss rate seems possible if
winds of A and hot F stars are 
radiative in nature. Since a star's \teff{} changes over time, 
a mass loss rate which depends on \teff{} 
(or on $L_*$) could also vary in time \citep{swenson92}. 
Such effects were not introduced in order to limit the 
number of adjustable parameters. 

The competition between atomic diffusion and meridional circulation in 2-D
should lead to solutions which resemble those obtained with mass loss,
since meridional circulation leads to an additional advective term 
in the transport equation (Eq.\,\ref{eqn:charb2}). Therefore, because the internal 
distribution of elements in the mass loss regime differs considerably from the
variations encountered in the turbulent mixing regime (see discussion in
Sect.\,\ref{sec:turb}), internal distributions due to meridional circulation could also 
differ significantly from those encountered via turbulence.
Hence, when building stellar models, one should be cautious when 
replacing meridional circulation, which is an advective process, by
turbulent mixing. 
A careful study of the atomic diffusion of metals 
within the context of meridional circulation, 
such as the one carried out for
helium in \citet{quievy09}, could help 
determine the implications of such an approximation. This could perhaps lead to 
asteroseismic tests which could distinguish between models using rotationally
induced turbulence (\citealt{talon06}) and those using meridional circulation
(\citealt{charbonneau88}), which are both used to explain the disappearance of the 
AmFm character for rotation velocities greater than 100\,km\,s$^{-1}$.  

Observations of rapid p-mode oscillations in many Ap stars 
(\citealt{kurtz78}) and in particular in Przybylski's star
(see also \citealt{mkrtichian08}) have led to a number of studies
of the oscillation mechanisms. In particular,
\citet{vauclair91} suggested that 
unseparated mass loss acting solely in polar regions,
where the magnetic field is strongest,
could induce helium gradients which are compatible with 
oscillation generating models
(\citealt{balmforth01}). Although 
differences between our stellar model and the 
one of \citet{vauclair91} could have an
effect on the predicted anomalies 
(notably the absence of convection due to magnetic 
braking/freezing in the latter),
our calculations suggest anomalies would 
reach inwards to about 25 \% of the star's radius
and could have an
important effect on opacities. Though it would depend on the
strength of the overall mass loss rate, which will be smaller than 
the mass loss rate at the poles, similar He depletions could
be coupled with overabundances of iron peak elements 
around 200\,000\,K and perhaps iron convection. 
It is not clear 
whether magnetic braking/freezing or thermohaline convection can stabilize
iron peak convection. Unfortunately, we are not able to investigate 
this scenario any further since 
our models require spherical symmetry.


Perhaps asteroseismology will allow us to answer some of these questions, while revealing the
relative importance of meridional circulation, turbulence and mass loss within
chemically peculiar stars.

\acknowledgements{We would like to thank G. Alecian as well as
the anonymous referee
for
useful comments which allowed us to improve
the paper. M.Vick thanks the D\'epartement de physique de 
l'Universit\'e de Montr\'eal for financial support, as well
as everyone at the GRAAL in Montpellier for their amazing hospitality. We acknowledge 
the financial support of Programme National de Physique 
Stellaire (PNPS) of CNRS/INSU, France. This research was
partially supported by NSERC at the Universit\'e de Montre\'al. Finally, we  
thank the R\'eseau qu\'eb\'ecois de calcul de haute performance (RQCHP) for providing us with the
computational resources required for this work.}

\clearpage

\end{document}